# Calibration-induced Systematics in SALT3 Training and Their Impact on Dark Energy Constraints from Stage IV Supernova Surveys

Kene Anumba[1], David O. Jones[2], Richard Kessler[3], Daniel Scolnic[1], W. D'Arcy Kenworthy[4], Rebecca C. Chen[5,6,7], Bastien Carreres[1], Maria Vincenzi[8], Erik R. Peterson[1,9], Maria Acevedo[1], Ben Rose[10], Dillon Brout[11], Jillian Paulin[12], Rujuta A. Purohit[13,14], and Rebekah Hounsell[15,16],
The Roman Supernova Cosmology Project Infrastructure Team
[1] Department of Physics, Duke University, Durham, NC 27708, USA
[2] Institute for Astronomy, University of Hawai'i, 640 N. Aohoku PI., Hilo, HI 96720, USA
[3] Kavli Institute for Cosmological Physics, University of Chicago, Chicago, IL 60637, USA
[4] Oskar Klein Center, Department of Astronomy, Stockholm University, SE-10691, Stockholm, Sweden
[5] Kavli Institute for Particle Astrophysics & Cosmology, PO Box 2450, Stanford University, Stanford, CA 94035, USA
[6] SLAC National Accelerator Laboratory, Menlo Park, CA 94025, USA
[7] Department of Physics, Stanford University, 382 Via Pueblo Mall, Stanford, CA 94305, USA
[8] Department of Physics, University of Oxford, Oxford OX1 3PU, UK
[9] Department of Physics, University of Michigan, Ann Arbor, MI 48109, USA
[10] Department of Physics and Astronomy, Baylor University, One Bear Place 97316, Waco, TX 76798-7316, USA
[11] Departments of Astronomy and Physics, Boston University, Boston, MA 02215, USA
[12] Department of Physics and Astronomy, University of Pennsylvania, 209 South 33rd Street, Philadelphia, PA 19104, USA
[13] Institute for Astrophysical Research, Boston University, 725 Commonwealth Avenue, Boston, MA 02215, USA
[14] Department of Astronomy, Boston University, 725 Commonwealth Avenue, Boston, MA 02215, USA
[15] UMBC, 1000 Hilltop Circle, Baltimore, MD 21250, USA
[16] NASA Goddard Space Flight Center, 8800 Greenbelt Road, Greenbelt, MD 20771, USA
*Received 2026 April 21; revised 2026 September 6; accepted 2026 September 8; published 2026 October 7*

## Abstract

In the coming years, the Vera Rubin Observatory's Legacy Survey of Space and Time (Rubin-LSST) and the Nancy Grace Roman Space Telescope's (Roman) High Latitude Time Domain Survey (HLTDS) are expected to discover more than a million Type Ia supernovae (SNe Ia), several orders of magnitude more than current samples and with a tighter control on systematic uncertainties. One of the largest systematic uncertainties in cosmological analyses with SNe Ia is the accuracy of the spectrophotometric model for SNe Ia time series data, which depends on the photometric calibration of the surveys. To quantify the impact of this uncertainty, we analyze simulated Rubin-LSST and HLTDS data, perturb the photometric zero-points and filter mean wavelengths, and propagate these systematics to spectral model recovery, estimated distances, and dark energy Figure of Merit (FoM) based on the $w_0w_a$CDM model. Zero-point shifts of 5 mmag and filter mean wavelength shifts of 5 Å lead to a ~50% decrease in the FoM relative to a statistical-only case when calibration uncertainties are propagated only through light-curve fitting. The same calibration shifts applied only during model training produce a smaller ~13% degradation. Contrary to previous analyses, calibration uncertainties in light-curve fitting dominate over those from model training. Their effect during light-curve fitting varies smoothly with redshift and is nearly degenerate with cosmology, preventing mitigation through self-calibration. Finally, we show that the FoM dependence on the size of the calibration uncertainties (in the range of expected sizes) is roughly linear.

## 1. Introduction

Over the past three decades, observations of Type Ia supernovae (SNe Ia) have been critical to both discovering the accelerating expansion of the Universe (A. G. Riess et al. 1998; B. P. Schmidt et al. 1998; S. Perlmutter et al. 1999) and improving measurements of dark energy properties that may be the source of the accelerated expansion. The results of these measurements (P. Astier et al. 2006; R. Kessler et al. 2009a; N. Suzuki et al. 2012; M. Betoule et al. 2014; D. M. Scolnic et al. 2018; S. Alam et al. 2021) have been largely consistent with a cosmological constant described by the dark energy equation of state parameter $w = -1$. More recent analyses of SNe Ia (D. Brout et al. 2022a; M. Vincenzi et al. 2024; D. Rubin et al. 2025), particularly when combined with other probes like the cosmic microwave background and baryon acoustic oscillation (Planck Collaboration et al. 2020; M. Abdul Karim et al. 2025; A. Adame et al. 2025), provide hints that the dark energy equation of state parameter may evolve with cosmic time. Despite the growing precision of cosmological measurements, the physical nature of dark energy remains unknown and constraining this property is one of the central goals of modern cosmology.

SNe Ia are standardizable distance indicators: despite an intrinsic luminosity scatter of ~0.8 mag, empirical correlations between peak luminosity, light-curve shape, and color (M. M. Phillips 1993; R. Tripp 1998) reduce this to ~0.1–0.17 mag (M. Hamuy et al. 1996; M. Betoule et al. 2014), enabling precise measurements of cosmic expansion history. In practice, this standardization is implemented using

**Table 1**
Key Characteristics of the Rubin-LSST and Roman SN Ia Surveys

| | Rubin-LSST | Roman |
|---|---|---|
| Telescope | Ground-based (VRO; Simonyi Telescope) | Space-based (L2) |
| Aperture | 8.0 m | 2.4 m |
| FOV | 9.6 $\deg^2$ | 0.28 $\deg^2$ |
| Filters | see Figure 1 | see Figure 1 |
| Spectroscopy | 4MOST & proposals | prism & grism[a] |
| Redshift | $0.025 < z < 1.1$ | $0.3 < z < 3.0$ |
| SN Ia count | $\sim 10^6$ | $\sim 1.5 \times 10^4$ |
| Cadence | WFD suboptimal, variable; DDF high, uniform | High, uniform |
| Primary references | M. Lochner et al. (2022), B. O. Sánchez et al. (2022) | D. Rubin et al. (2025), H. HLTDS (2025) |

**Note.**
[a] Only prism spectra are generated in our simulation.

spectral time-series models such as SALT (J. Guy et al. 2005, 2007), which parameterize multicolor light curves as a function of phase, wavelength, shape, and color. W. D. Kenworthy et al. 2021, hereafter K21 introduced SALT3, an upgraded SALT model, and released SALTShaker, a publicly available `Python`-based training pipeline with improved uncertainty handling and color–stretch separation, significantly improving the incorporation of photometric calibration uncertainties into the training process (G. Taylor et al. 2023; B. Popovic et al. 2026). This pipeline has since been applied to extend SALT into the near-infrared (J. D. R. Pierel et al. 2022), study training sample systematics (M. Dai et al. 2023), measure host-galaxy mass dependence of the SN Ia spectral energy distribution (SED; D. O. Jones et al. 2023; G. Taylor et al. 2024), assess additional principal components (W. D. Kenworthy et al. 2025), and probe ultraviolet redshift evolution (Q. Wang et al. 2025). Nevertheless, the standardization process above relies on accurate modeling of the supernova SED and precise photometric calibration and filter characterization. Uncertainties in these quantities propagate directly into distance estimates and can bias inferred cosmological parameters, making them one of the dominant sources of systematic uncertainty in current analyses (M. Betoule et al. 2014; D. Scolnic et al. 2014; D. Brout et al. 2019, 2022a; D. O. Jones et al. 2019; M. Vincenzi et al. 2024; B. Popovic et al. 2026).

Addressing these limitations to improve constraints on dark energy requires both larger SN Ia samples and tighter control of systematic uncertainties. The upcoming Vera Rubin Observatory's Legacy Survey of Space and Time (Rubin-LSST; see LSST Science Collaboration et al. 2009; Z. Ivezic et al. 2019) and the Nancy Grace Roman Space Telescope's (Roman) High Latitude Time Domain Survey (HLTDS; see D. Spergel et al. 2015; R. Hounsell et al. 2018; B. M. Rose et al. 2021) are examples of Stage IV dark energy experiments, defined by A. Albrecht et al. (2006) as major long-term programs that improve constraints on dark energy by at least a factor of 10 in the Dark Energy Task Force's (DETF) figure of merit and a factor of 3 on the uncertainties in $w_0$ and $w_a$ relative to Stage II surveys (e.g., SDSS, SNLS). See some characteristics of the surveys in Table 1. Together, both surveys will provide unprecedented SNe Ia measurements across complementary redshift ranges, increasing sample sizes by more than 2 orders of magnitude. With such large statistics, controlling systematic uncertainties becomes essential to fully realize the cosmological potential of these surveys.

In this paper, we perform a cosmology analysis on simulated data of Rubin-LSST and HLTDS with the goal of forecasting the impact of photometric calibration uncertainties. Key contributors to these systematics are uncertainties in zero-points and filter transmissions (see Section 6.1 of R. Hounsell et al. 2018). These uncertainties propagate into cosmological parameter estimation through two pathways: first, during the training of the SALT3 model, and second, through their effect on the light-curve fits produced by this model. D. Brout et al. (2022b, Section 5.2) showed that propagating these uncertainties both at the SALT model training and at light-curve fitting simultaneously produces roughly twice the change in the recovered cosmology compared to modifying either component on its own. We then trace the impact of these uncertainties all the way through to constraints on the equation of state of dark energy, parameterized by $w_0$ and its possible evolution $w_a$.

Within the Roman Supernova Project Infrastructure Team (SNPIT), this work continues the effort to analyze HLTDS simulations and optimize the dark energy figure of merit (FoM). Building on R. Kessler et al. (2055, hereafter K25), who used covariance matrices to quantify the impact of multiple systematic uncertainties on the FoM, we extend the analysis with key refinements. The first major difference is that K25 used the same SALT3 model for the entire cosmology pipeline, whereas in this study we simulate the data with the same input model as K25 but adopt a different procedure by using these data to train a new SALT3 model for subsequent pipeline stages. Also, in anticipation of including higher-redshift spectra from the Roman prism to model potential redshift-dependent effects in the SALT3 model, here we use only prism spectra in the training to further test SALTShaker (see B. A. Joshi et al. 2022, for more details on prism spectra). Second, while K25 propagated calibration systematics only during the light-curve fitting stage, we additionally propagate these systematics through the SALT3 model training itself. While K25 made use of photometric SN +host redshifts, here we adopt a simplified and idealized scenario in which all events are assigned spectroscopic redshifts (see R. C. Chen et al. 2026 for more details on HLTDS spectroscopic redshifts) from their host galaxies, reducing both analysis complexity and computational cost. Further, we expect the impact from this simplification to be small as previous forecasts and data-based comparisons have shown that the difference is modest (M. Dai et al. 2018; R. Chen et al. 2022; A. Mitra et al. 2023; R. C. Chen et al. 2025). Our approach is similar to analyses by M. Betoule et al. (2014), D. Brout et al. (2022a), M. Vincenzi et al. (2024), and B. Popovic et al. (2026), who quantified calibration-related impacts by propagating uncertainties through both model training and light-curve fitting using observed data rather than simulations. D. Rubin et al. (2025) also studied calibration-driven impacts on parameter estimation using a Fisher-matrix approach. However, many previous cosmological forecasts, such as R. Hounsell et al. (2018), did not include the training of the light-curve model within the systematic uncertainty budgets. Forecasting the impact of calibration systematics, the focus of this work, is crucial to undertake before surveys begin, as it can inform survey strategy and help define requirements on calibration accuracy, filter characterization, and related systematics needed to mitigate their impact.

The structure of this paper is as follows. Section 2 describes our methodology, light-curve simulation and fitting, model training, and cosmological parameter estimation. Section 3 presents the results of the model training, the different analysis variants and the impact of systematic uncertainties. We discuss and conclude the paper in Section 4.

## 2. Methodology

Below, we summarize the analysis steps used to understand the impact of calibration uncertainties on the dark energy FoM. We elaborate on these procedures in Sections 2.1–2.6.

1. Simulation: Simulate realistic light-curve samples for Rubin-LSST and Roman using the filter transmission curves for both surveys, along with Roman prism spectra to use in the SALT3 training. We use an optical+NIR SALT model, `SALT3.NIREXT`, as the true input SED model to generate all light curves and spectra. In addition to the SN Ia sample, we also simulate non-SN Ia events like core-collapse supernovae to represent the contaminants and include spectroscopic redshifts from host galaxies (zSpec).
2. Training: Train the SED model on a high-quality SN Ia subset to obtain a nominal model as well as "biased" models that are produced using shifted zero-points and filter transmissions.
3. Light-curve fitting: Fit the light curves from the full sample in step (1) with all of the SED models from step (2).
4. Classification: Run the SCONE classifier (H. Qu et al. 2021). SCONE is an SN classifier used to determine the probability of each SN candidate being of Type Ia. These probabilities are used to weight each event during the distance modulus measurement stage.
5. Distance moduli: Convert light-curve parameters to standardized distance moduli, and correct distances for biases from selection effects and non-SN Ia contamination.
6. Systematic accounting: For each calibration shift in step (2), compare recovered, biased distance modulus values to those recovered when light curves are fit with the nominal model, and propagate differences into a systematic covariance matrix.
7. Cosmological inference: Propagate the systematic covariance matrix to a cosmology fit to constrain cosmological parameters. Compare constraints to those not including systematic uncertainties.

### 2.1. Simulation

In this section, we describe the simulation framework used to model the survey conditions for Rubin-LSST and HLTDS. These simulations provide the foundation for both SALT3 model training and subsequent cosmological analysis variants.

#### 2.1.1. Survey Simulation Framework

We simulate Rubin-LSST and HLTDS following K25 with the SuperNova ANAlysis software package (SNANA; R. Kessler et al. 2009b).

SNANA generates catalog-level simulations of transients that realistically reflect the predicted survey conditions, including observational effects such as depth per filter and cadence (see R. Kessler et al. 2019a; J. D. R. Pierel et al. 2021; A. C. Bailey et al. 2023 for more details on SNANA-based survey simulations).

The LSST observation sequence is simulated to match the Baseline v3.4 results of the Operations Simulator (F. Delgado et al. 2006; M. A. Reuter et al. 2016). Our simulated observations include both components of the 10 yr Rubin-LSST survey: the wide–fast–deep (WFD) fields, which probe low-redshift SNe, and the deep drilling fields (DDF), which extend to higher redshifts. This survey strategy prioritizes stripes of the sky for repeat observation in a "rolling" cadence over a period of several years to boost transient discovery and light-curve sampling. The LSST passbands (*ugrizY*) cover optical wavelengths ($3000 < \lambda < 11000$ Å), see Figure 1.

The HLTDS will span a total observing time of around six months, distributed over a two-year period. The observing cadence varies by component: the bluest bands and the prism are observed at roughly a 5 day cadence, while the remaining bands follow a 10 day cadence across the shallow and deep surveys. Presurvey imaging and prism spectroscopy will provide photometric redshifts and spectroscopic redshifts for a subset of host galaxies, as well as templates for difference imaging (see Figure 1 and Table 1 of K25). Additional spectroscopic observations from external facilities (e.g., Subaru) are expected. Roman filters cover the optical-NIR ($6000 < \lambda < 20{,}000$ Å). For a subset of SNe Ia, spectra will be obtained using the prism, a dispersive element designed for slitless, multiobject spectroscopy, which covers a wavelength range from 7500 to 18000 Å with a resolution of about 100 (see Section 2.3 of B. M. Rose et al. 2025).

#### 2.1.2. Generated Samples

For this study, we generate a single simulated dataset for each survey, from which we select a high-quality subset to train the SALT3 model (see Section 2.2). The Roman training set additionally includes spectra obtained from the prism. The prism-to-event ratio was selected empirically to ensure stable SED model training: Roman's higher signal-to-noise and NIR spectral information require a larger simulated sample to appropriately weight its contribution relative to Rubin-LSST. The full simulation is used for cosmology analysis; in these cosmology samples, Roman events also include core-collapse supernovae and other contaminants like peculiar SNe Ia (e.g., 91bg-like and Iax-like objects). The simulated SN distribution across redshift is shown in Figure 2. "All sim data" represents the full simulation used for cosmology analysis, including contaminants in the Roman sample, whereas "Training data" is a high-quality subset of this simulation used for SALT3 model training. For the simulations, the stretch and color populations are from the dust model in B. Popovic et al. (2023). Furthermore, we adopt the same volumetric rate model as PLAsTiCC (R. Kessler et al. 2019b) to estimate the expected number of supernovae within a given volume and time frame: see B. Dilday et al. (2008) and R. Hounsell et al. (2018) and Equations (1)–(2) in R. Kessler et al. (2019a) for more details on volumetric rates.

After applying a series of quality cuts, the training sample includes 1039 Rubin-LSST and 2491 Roman SN light curves, together with 4388 prism spectra obtained for a subset of

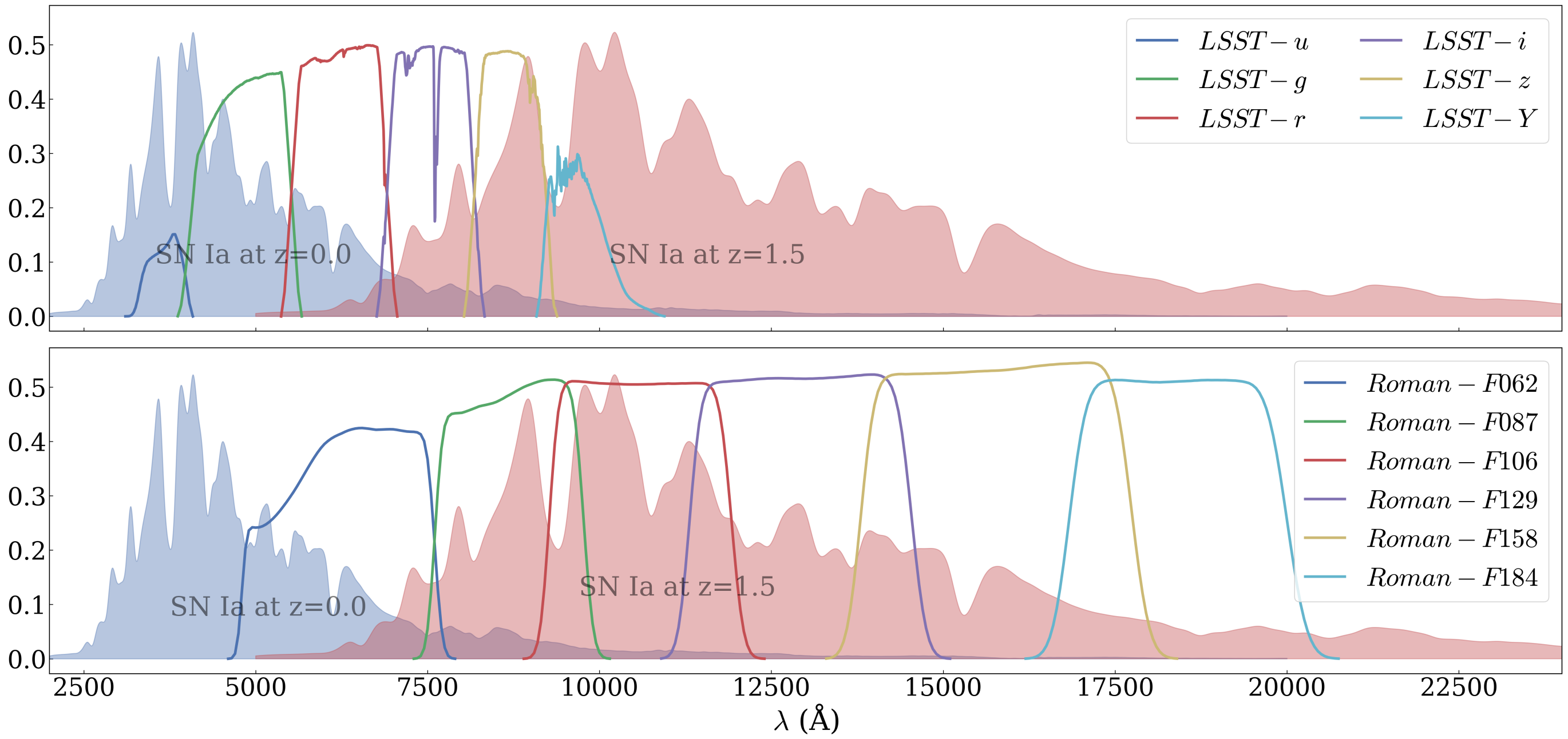

**Figure 1.** Comparison of LSST and Roman Space Telescope filter transmission curves overlaid on SN Ia spectra at low and high redshift. The figure illustrates the wavelength coverage of each filter set and how they sample the SN Ia spectra at different redshifts, highlighting Roman's extended near-infrared sensitivity, which is essential for observing high-redshift SNe Ia.

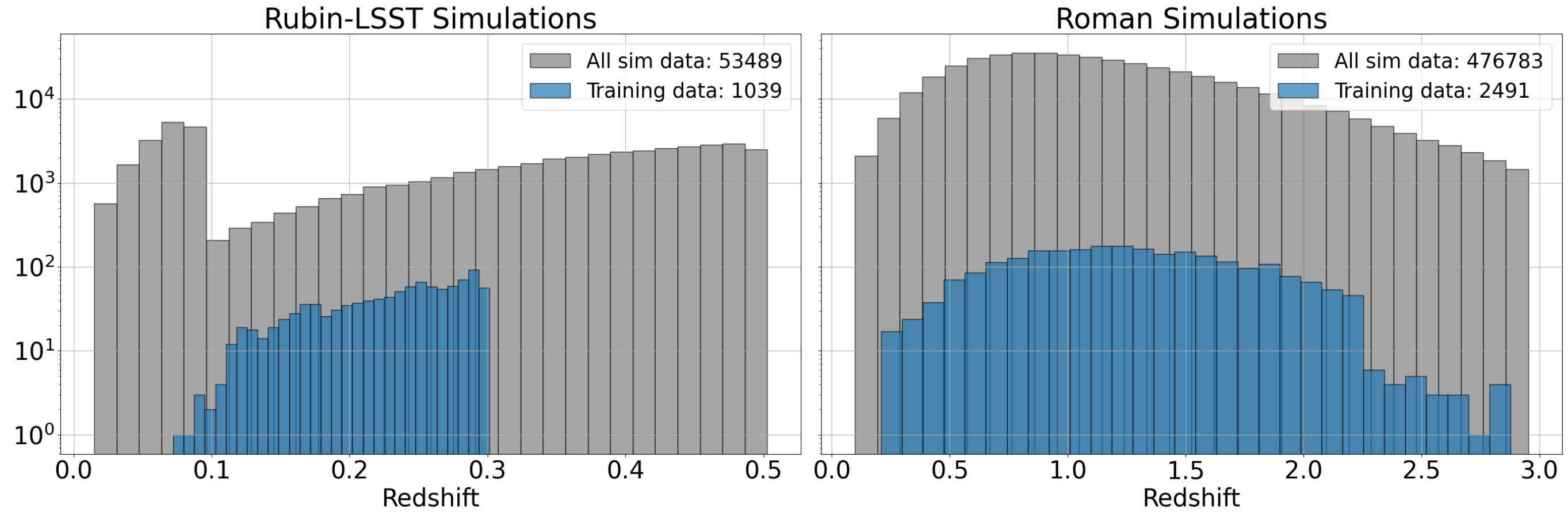

**Figure 2.** The gray-colored distributions illustrate the number of SNe Ia with respect to redshift that we simulated for both LSST (left) and Roman (right). The $y$-axis is in log scale. To ensure high-quality simulated data for training, we applied several selection cuts on the light curves and spectra. The final training set consisted of 1039 LSST SNe, 2491 Roman SNe and 4388 Roman prism spectra. Our chosen redshift ranges for both surveys are designed to align with the core community survey (Roman Observations Time Allocation Committee & Core Community Survey Definition Committees 2025) recommendations and minimize complications arising from LSST selection effects above redshifts of 0.3.

Roman events, with some supernovae contributing multiple spectra. The requirements that each SN must satisfy are:

1. At least four total photometric epochs between −10 to +35 days relative to the time of peak brightness.
2. At least one observation within −10 to +5 days of peak brightness and at least one within +5 to +20 days.
3. Photometric coverage in at least three filters, each with at least one observation within −8 to +10 days of peak brightness.
4. For the prism spectra (training only), the average signal-to-noise ratio (SNR) is greater than 10 over rest-frame wavelength range 2500–11,000Å.
5. For LSST SNe Ia in the training sample, we included only those with $z < 0.3$, following core community survey recommendations (Roman Observations Time Allocation Committee & Core Community Survey Definition Committees 2025) and limiting the impact of selection effects at higher redshift.

A representation of the light-curve and spectral coverage in wavelength and phase space can be found in Figure 3.

For cosmology analysis, we generate and analyze nine statistically independent datasets for both surveys, using `Pippin` (S. Hinton & D. Brout 2020) to orchestrate the different SNANA stages. Each Roman simulation dataset includes SNe Ia, peculiar SNe Ia (Iax and Ia-91bg), and core-collapse supernovae such as IIP, IIL, Ib, Ic (see K25 for details on the source models for these simulations). The only selection applied to these samples is the "common event" requirement (see Section 2.6), which retains only supernovae that are unaffected by systematic-driven cuts. Furthermore, we generate larger "biascor" samples that are used for classifier probability prediction, bias corrections, and non-SNIa prior in BBC (see Section 2.4 below). Because of the large biascor size, we generate only one biascor sample and apply it to all nine datasets.

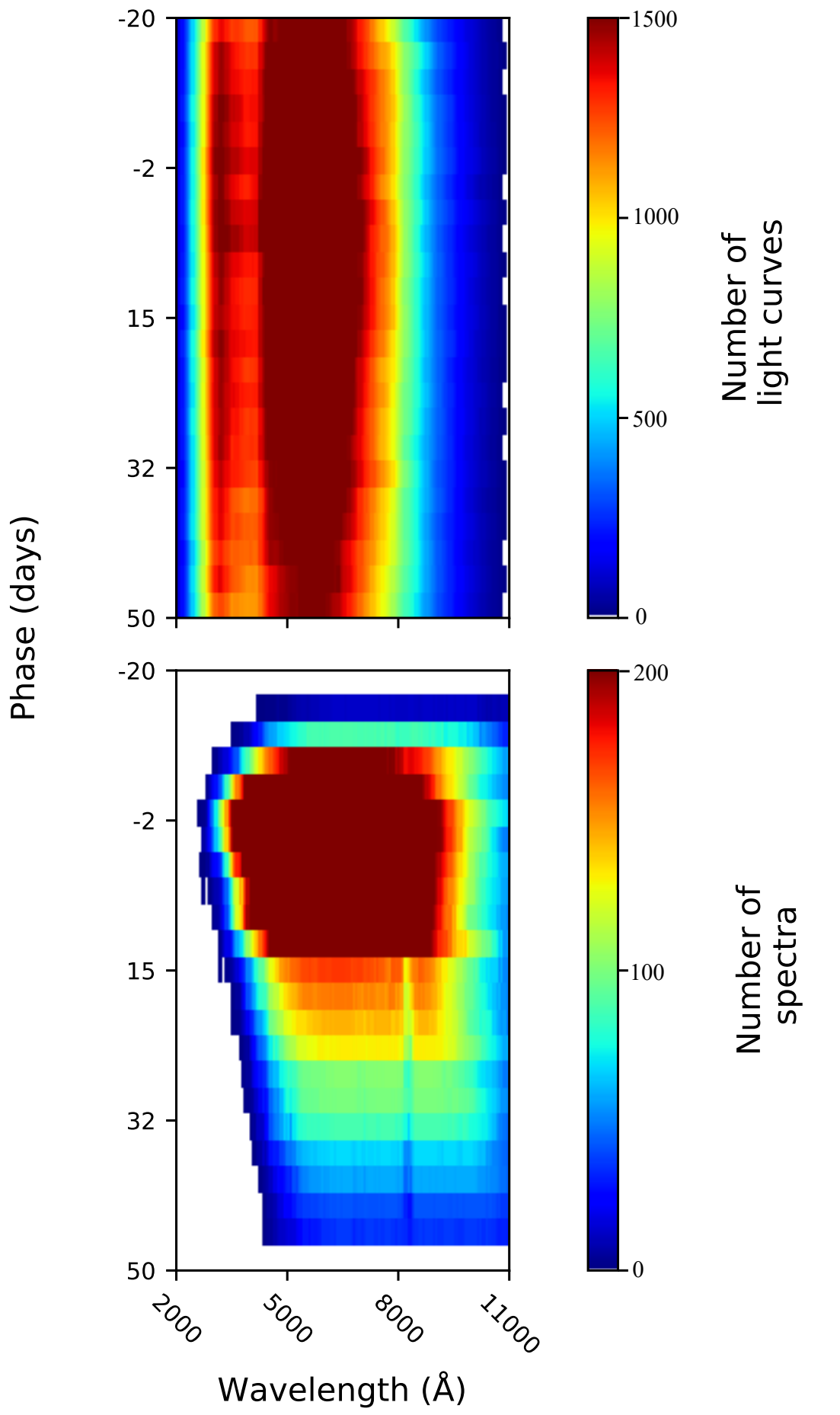

**Figure 3.** Number of light curves from all surveys used in model training to constrain each phase/wavelength bin. Photometric and spectral coverage are shown in the top and bottom panels, respectively. For each plot, the broad lines are due to filter transitions.

## 2.2. Training

### 2.2.1. Input Model

As described in Section 1, the SALT3 model flux density $F(p, \lambda)$ is parametrized as:

$$F(p,\lambda) = x_0[M_0(p,\lambda;\boldsymbol{m_0}) + x_1 M_1(p,\lambda;\boldsymbol{m_1})] \cdot \exp(c \cdot CL(\lambda;\boldsymbol{cl})), \quad (1)$$

where $M_0(p, \lambda;\boldsymbol{m_0})$ and $M_1(p, \lambda;\boldsymbol{m_1})$ are flux surfaces, analogous to principal components, representing the SED of a fiducial SN Ia and a first-order correction, respectively. $CL(\lambda;\boldsymbol{cl})$, is a color law that incorporates both the effects of intrinsic color variation and host-galaxy dust extinction (see the parameterization equation in Section 2 of K21). The $x_0$ (the overall flux normalization), $x_1$ (shape), and $c$ (color) parameters are unique to each SN. The $\boldsymbol{m_0}$ and $\boldsymbol{m_1}$ parameters are the set of second-order B-spline knots that define the model surfaces.

K21 define the SN Ia spectral energy distribution over a wavelength range of 2000 to 11000 Å and rest-frame phase range of −20 to +50 days relative to peak brightness. As described in Section 1, P22 extended this to 20000 Å, and for the OpenUniverse2024 image simulation project (OpenUniverse et al. 2024) it was further extended to 25000 Å using the BYOSED (J. D. R. Pierel et al. 2021) tool. The resulting `SALT3.NIREXT` model serves as the input (true) model for generating prism spectra and simulated light curves used in this analysis.

### 2.2.2. Model Training and Recovery

*SALTShaker* takes as input the simulated light curves and spectra, the initial estimates for the time of peak brightness, the SN parameters ($x_0$, $x_1$, and $c$, which are determined by doing a fit for these parameters using the SNANA fitting package), and the model parameters ($\boldsymbol{m_0}$, $\boldsymbol{m_1}$, $\boldsymbol{cl}$; see Section 2 of K21 for more details). The output of this process is a set of trained model templates.

To validate the trained models, we compare the recovered spectral surfaces to those of the true input model, `SALT3.NIREXT`, in Figure 4. The UV region shows larger discrepancies between the trained and input models. This discrepancy is primarily due to limited availability of prism spectra in this wavelength range, which reduces the model's ability to constrain the UV behavior. We further compare the true versus trained model performance at recovering SN parameters for simulated Roman light curves in Appendix A; both models show broadly similar distributions for most of the SN parameters and their uncertainties.

## 2.3. Light-curve Fitting and Standardization

We apply SALT3 light-curve fitting at two stages in this work. The first is applied to the training sample to obtain initial estimates of the SN light-curve parameters: stretch ($x_1$), color ($c$), time of peak brightness ($t_0$), and amplitude ($x_0$), required to initialize *SALTShaker* model training. Since this is a simulation study where the true underlying model is known, we use `SALT3.NIREXT` for these fits; in a real survey analysis, a previously trained model would serve this role. The second stage fits the cosmology analysis sample with our trained nominal and calibration-biased models to extract standardization parameters for distance measurements and cosmological inference.

We use the SNANA framework to fit light-curve parameters $x_0$, $x_1$, $c$, $t_0$, as introduced in Equation (1). We use the Tripp estimator (R. Tripp 1998), with modifications from R. Kessler & D. Scolnic (2017), to convert the light-curve parameters to distance modulus values using:

$$\mu = m_B + \alpha x_1 - \beta c - \Delta_B + \gamma - M_B. \quad (2)$$

The nuisance parameters $\{\alpha, \beta, \gamma, M_B\}$ are constrained from the SN sample in the BBC fit (see Section 2.4.1) where $\alpha$, $\beta$ are luminosity–stretch and luminosity–color correlation parameters, respectively, $\gamma$ is a correction for the host-mass correlation (P. L. Kelly et al. 2010; H. Lampeitl et al. 2010; M. Sullivan et al. 2010), and $M_B$ is the SN Ia absolute magnitude. $\Delta_B$ is a bias correction term based on simulations (see Section 2.4 below). $m_B$, the synthetic apparent magnitude in the rest-frame $B$-band, is related to $x_0$ through the relation:

$$m_B = -2.5 \log_{10}(x_0) + \text{constant}. \tag{3}$$

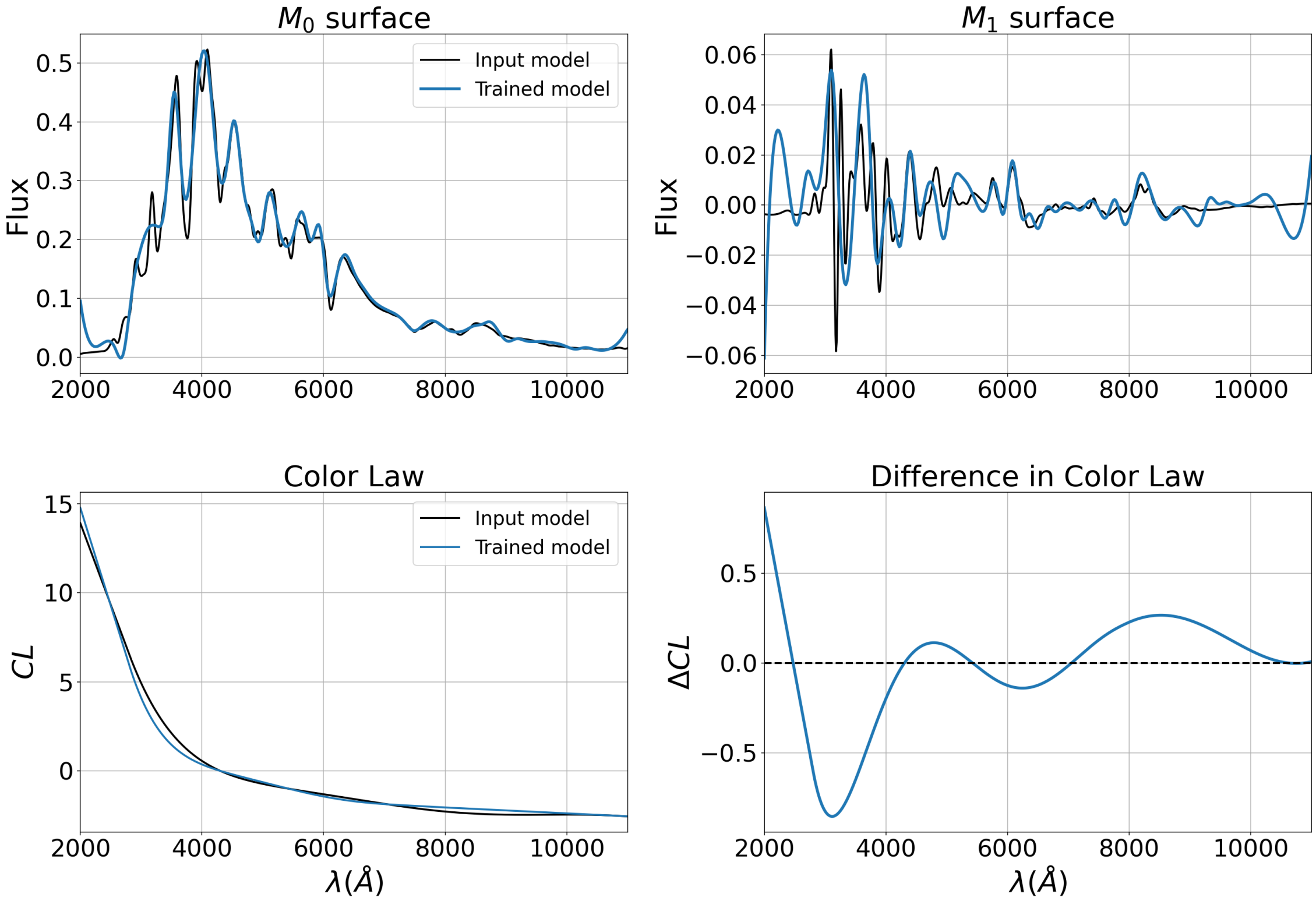

**Figure 4.** The top panels illustrate the accuracy of recovering the flux surfaces, $M_0$ (left) and $M_1$ (right), which correspond to the spectral energy distribution (SED) of a fiducial SN Ia and its first-order correction, respectively, while the bottom panels show the color law (left) and its difference relative to the input model, `SALT3.NIREXT` (right).

### 2.4. Bias Correction, Contamination Mitigation, and Systematic Treatment

#### 2.4.1. BEAMS with Bias Corrections (BBC Formalism)

Only a small subset of Roman transients will include spectroscopic classification, and therefore our simulated sample includes non-Ia contamination from core-collapse SNe (II/Ib/Ic) and peculiar SNe Ia (Iax, 91bg). To simplify the analysis, we do not include contamination in the Rubin-LSST samples.

The presence of non-Ia contaminants in photometric samples can bias distance measurements and cosmological inference. Bayesian Estimation Applied to Multiple Species (BEAMS; M. Kunz et al. 2007; R. Hlozek et al. 2012; D. O. Jones et al. 2018, 2019; R. Kessler et al. 2023) addresses this by modeling mixed samples and incorporating photometric classification probabilities. In our analysis, we use BEAMS with Bias Corrections (BBC, J. Marriner et al. 2011; R. Kessler & D. Scolnic 2017) to implement the BEAMS formalism described above, and we use the SCONE classifier from H. Qu et al. (2021) to obtain the prior probability that a given transient is an SN Ia. BBC uses a large simulation to apply distance bias corrections to each event. BBC is implemented by the `SNANA` program `SALT2mu` (J. Marriner et al. 2011), which performs a global fit to the data to determine distances that are corrected for selection effects and non-Ia contamination. The BBC fit also determines the nuisance parameters $\alpha$, $\beta$, $\gamma$ and intrinsic scatter $\sigma_{\text{int}}$. BBC-corrected distances are used to construct the Hubble diagram (HD).

#### 2.4.2. Covariance Matrix

BBC distances are computed both for the baseline analysis and for a suite of variant analyses, which are used to quantify systematic uncertainties and to construct a statistical+systematics covariance matrix. Following the formalism of A. Conley et al. (2011) and D. Brout et al. (2019, see Section 3.8.2 of the study), we compute the SN uncertainty covariance matrix $\mathcal{C}_{\text{tot}}$, as a sum of a diagonal statistical component ($\mathcal{C}_{\text{stat}}$) and a systematic component ($\mathcal{C}_{\text{syst}}$);

$$\mathcal{C}_{\text{tot}} = \mathcal{C}_{\text{stat}} + \mathcal{C}_{\text{syst}}. \tag{4}$$

$\mathcal{C}_{\text{tot}}$ is typically an $N \times N$ matrix, where $N$ denotes the number of SNe in the HD. To reduce CPU resources from large matrices, particularly in the cosmology fitting stage, we use the REBIN approach introduced in R. Kessler et al. (2023), where $N$ corresponds to the number of 3D bins in redshift (45), stretch (4), and color (8), resulting in $N = 1440$. A small fraction of these 3D bins contain no events, and therefore the effective HD size is around 1300.

$C_{\text{stat}}$ is a diagonal matrix whose $i$th entry is the BBC-fitted $\mu$-uncertainty in the $i$th bin in redshift, stretch, and color. $C_{\text{syst}}$ is computed from some of the systematic uncertainties $S$, summarized in Table 2. Following M. Vincenzi et al. (2024),

**Table 2**
Nominal Systematic Uncertainties Used for Covariance Matrix Calculation

| Systematic | Description | Value(s) |
|---|---|---|
| Cal_nonlin | Nonlin in Roman detectors | ∼0.98 - 1 |
| Cal_per_A | Global calibration | 0.0071 |
| Cal_SALT3 | Filter zero-point & $\lambda$ shift[a] | 5 mmag, 5 Å |

**Note.**
[a] Included in SALT3 training and/or light-curve fit.

we compute the systematic covariance matrix using:

$$C^{ij}_{syst} = \sum_{S=1}^{N_S} \langle \Delta\mu^i_S \rangle \langle \Delta\mu^j_S \rangle W^2_S, \tag{5}$$

where $W_S$ is the weight associated with each systematic $S$ ($W_S = 1$ unless specified otherwise) and the indices $i$ and $j$ run over the total number of bins as described above. Defining $\mu NOMi$ as the nominal distances and $\mu Si$ as the BBC-fitted distances for systematic $S$, we define distance shifts relative to our nominal analysis

$$\langle \Delta\mu^i_S \rangle \equiv \langle \mu^i_S \rangle - \langle \mu^i_{\rm NOM} \rangle. \tag{6}$$

Roughly 2% of the simulated SNe are excluded due to a common event requirement (Section 2.6), in which selection cuts and a valid bias correction (Section 2.4.1) are required for each of the systematic realizations.

### 2.5. Cosmological Parameter Inference

Each dataset is analyzed to constrain the cosmological parameters $\{\Omega_M, w_0, w_a\}$ within the $w_0 w_a$CDM framework. The fit is performed by minimizing the following chi-squared function:

$$\chi^2 = \Delta\mu^T \mathcal{C}^{-1}_{\rm tot} \Delta\mu + \chi^2_{\rm prior}, \tag{7}$$

where $\Delta\mu = \mu_{\rm BBC} - \mu_{\rm theory}(z;\Omega_M, w_0, w_a)$.

Parameter estimation is carried out using SNANA's grid-based likelihood search, which evaluates the likelihood over a three-dimensional grid ($\Omega_M$, $w_0$, $w_a$) and returns marginalized constraints along with their uncertainties.

After obtaining the fit results, we follow A. Albrecht et al. (2006) and Y. Wang (2008) to compute the Dark Energy Task Force figure of merit (FoM) as:

$$\mathrm{FoM} = [\sigma_{w_0}\sigma_{w_a}\sqrt{1 - \rho^2}]^{-1}, \tag{8}$$

where $\sigma_{w_0}$ and $\sigma_{w_a}$ are the marginalized uncertainties on $w_0$ and $w_a$, respectively, and $\rho$ is the reduced covariance between these two parameters. The FoM is used to quantify the constraining power on dark energy properties. The larger the value of the FoM, the tighter the constraints.

We define $\langle \mathrm{FoM} \rangle$ for each analysis variant (see Section 3.1) as an average of the FoMs of all systematic realizations in that analysis variant.

$$\langle \mathrm{FoM} \rangle = \frac{1}{N}\sum_{i=1}^{N} \mathrm{FoM}_i\,, \tag{9}$$

where $N$ is the number of systematic realizations and $\mathrm{FoM}_i$ is the figure of merit for the $i$-th realization.

**Table 3**
Average Event Statistics Showing Baseline Event Counts and the Fraction Retained after Common Events Loss for Train+Fit (LR) Fixed and Train+Fit (LR) Random

| Sample | Baseline | Fixed | Random |
|---|---|---|---|
| ALL | 14650.8 | 0.990 | 0.987 |
| LSST (WFD) | 595.6 | 0.942 | 0.931 |
| LSST (DDF) | 3630.7 | 0.984 | 0.980 |
| Roman (WIDE) | 6427.4 | 0.996 | 0.994 |
| Roman (DEEP) | 3997.1 | 0.993 | 0.990 |

**Table 4**
Statistical $\mathcal{R}_{\rm FoM}$ for Different Analysis Configurations (Section 3.1)

| Configuration | $\mathcal{R}_{\rm FoM}$ |
|---|---|
| Train (LR) Fixed | 0.977 |
| Fit (LR) Fixed | 0.998 |
| Train+Fit (LR) Fixed | 0.989 |
| Train (R) Random | 0.974 |
| Fit (R) Random | 0.987 |
| Train+Fit (R) Random | 0.984 |
| Train (L) Random | 0.971 |
| Fit (L) Random | 0.992 |
| Train+Fit (L) Random | 0.984 |
| Train (LR) Random | 0.971 |
| Fit (LR) Random | 0.979 |
| Train+Fit (LR) Random | 0.982 |

**Note.** The Small Degradation Arises from Common Event Loss.

We also define a baseline analysis in which no systematics are applied (hence stat-only), and there is no event loss from the common event requirement (Section 2.6). This baseline analysis provides the optimal constraint with

$$\mathrm{FoM}_{\rm baseline} \simeq 615. \tag{10}$$

In this study, the metric for FoM degradation is described by the ratio

$$\mathcal{R}_{\rm FoM} \equiv \langle \mathrm{FoM} \rangle / \mathrm{FoM}_{\rm baseline}. \tag{11}$$

### 2.6. Impact of Common Events Loss

Photometric calibration systematics in the light-curve fitting can induce unphysically large changes in SALT3 parameters, causing some SNe Ia to fail quality cuts that they would otherwise pass or acquire uncertainties that effectively exclude them from the cosmology fit (see Section III of A. Mitra et al. 2026). As A. Mitra et al. (2026) notes, this occurs rarely for any single systematic, the effect compounds across multiple systematics, leading to a nonnegligible cumulative sample loss. Given the focus of our paper and to ensure a fair comparison of FoM values across analysis variants (Section 3.1), we retain only those events unaffected by these systematic-driven cuts, a selection we refer to as the "common event" criterion.

While analysis variants with systematics will have $\mathcal{R}_{\rm FoM} < 1$, the stat-only analysis variants in this section also have $\mathcal{R}_{\rm FoM} < 1$ because of the common event criterion that leads to this degradation, an example of such cut is shown in Table 3. Table 4 shows the impact of common event criterion on the FoM degradation across the different analysis variants (see Section 3.1).

## 3. Impact of Calibration Systematics

### *3.1. Overview*

To assess the impact of photometric calibration systematics on light-curve model training and cosmological inference, we conduct a series of controlled analysis variants. We organize our systematic tests into three sets of variants:

1. Survey: LSST WFD+DDF (L), Roman WIDE+DEEP (R), LSST+Roman (LR)
2. Systematics: TRAIN, FIT, TRAIN, FIT
3. Perturbation: shift each filter individually (FIXED), or random shift drawn from a normal distribution for all filters (RANDOM)

For the Survey variant, we include either one sample, or both SN samples, For the Systematic variant, TRAIN refers to propagating calibration systematics in the data used for training, but not in data used for light-curve fitting and cosmological inference; FIT refers to propagating calibration systematics in the data used for light-curve fitting, but not for data used in the training; TRAIN, FIT refers to propagating systematics in the data used for both training and light-curve fitting. For the Perturbation variant, FIXED refers to applying fixed zero-point shifts individually for each filter, and similarly for the filter-wavelength shifts; this results in 12 systematic contributions for L, 12 for R, and 24 for LR. RANDOM refers to applying random-normal shifts to all bands for each of the 30 separate SALT3 trainings.

We choose calibration systematic uncertainty values to be 5 mmag for zero-point offsets and 5 Å for central filter wavelength. While LSST and Roman are aiming to improve upon this systematic, these values reflect the precision achieved so far by ground-based imaging surveys (D. Brout et al. 2022b; B. Popovic et al. 2026). For the FIXED method, each systematic contribution includes one systematic shift. For RANDOM, we apply Gaussian-distributed zero-point offsets (mean 0, $\sigma = 5$ mmag) and central wavelength shifts (mean 0, $\sigma = 5$ Å). We also explore the FoM dependence from changing the nominal systematics values.

For the RANDOM method, we investigate how the number of model realizations influences the FoM in Appendix B. While most of the current literature typically adopts nine realizations (M. Vincenzi et al. 2024; B. Popovic et al. 2026), we test configurations with 4, 9, 20, and 30 to 100 (in increments of 10) realizations, properly weighting the uncertainties in each case. These tests are identical to the TRAIN, FIT (LR) setup, varying only in the number of realizations used. From these tests, we find that the FoM is effectively converged by ∼60 realizations. The analysis variants presented here use 30 realizations, where the FoM is already close to convergence, as doubling the number of realizations for all variants would incur a significant computational cost. Based on this study, the FoM degradation reported in this paper may be overestimated by approximately ∼10% (see Appendix B).

### *3.2. Impact on Light-curve Model and Distance Modulus Estimation*

We show in Figure 5 how calibration uncertainties affect SALT3 flux surfaces as well as the relative difference in the flux surface. The uncertainties affect not only the principal flux surface ($M_0$) but also the higher-order component ($M_1$), altering the full spectral energy distribution used to fit SNe.

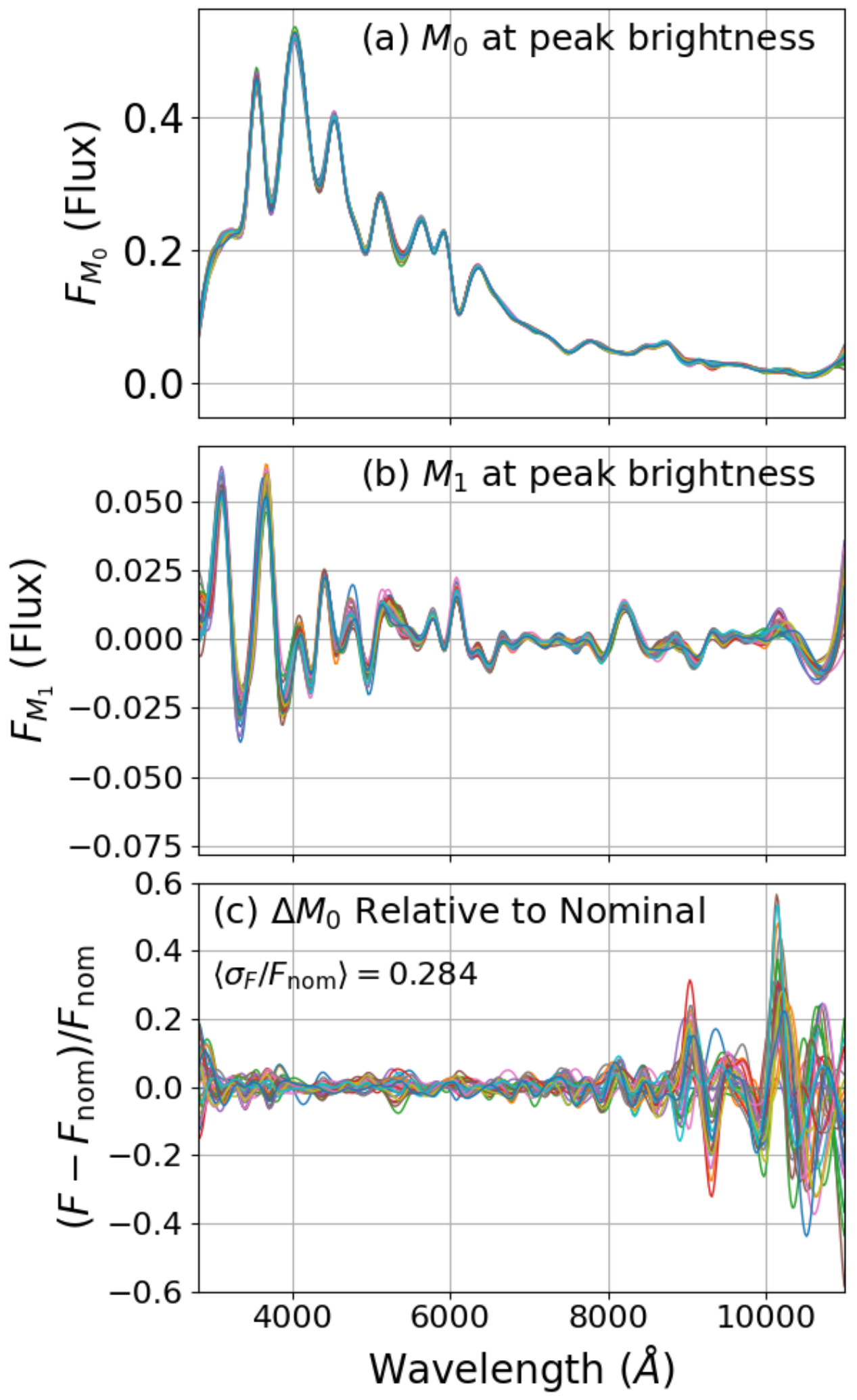

**Figure 5.** Comparison of trained model components illustrating the effect of random calibration variations shown in different colors. Panels (a) and (b) are $M_0$ (mean spectral component) and $M_1$ (first variability component) at peak brightness for different calibration realizations. (c) Fractional change in $M_0$ relative to the nominal model. $F_{\rm nom}$ refers to the flux for the surface without any systematics. $\langle \sigma_F/F_{\rm nom} \rangle$ gives the mean fractional standard deviation of $M_0$ across all calibration variants, quantifying the typical amplitude of calibration-induced flux perturbations.

The mean fractional standard deviation across calibration variants given in the figure, $\langle \sigma_F/F_{\rm nom} \rangle$, quantifies the typical amplitude of these perturbations across the full wavelength range.

### *3.3. Impact on Cosmology Constraints and Figure of Merit (FoM)*

We show in Figure 6 the redshift trend of the change in BBC-fitted distance modulus residuals with respect to the baseline analysis, for calibration uncertainties (Table 2) applied to TRAIN, FIT, and TRAIN, FIT. We define the $y$-axis label of Figure 6 as:

$$\Delta\mu_{\rm syst} = \mu_{\rm syst} - \mu_{\rm baseline}. \quad (12)$$

Our simulation of the two surveys overlaps around a redshift of 0.5, and this overlap causes the noticeable bumps in the scatter amplitude in the panels, especially in the FIT (middle) column. Furthermore, the training procedure introduces noticeable wiggles in the Train (left) and Train, Fit (right) columns of Figure 6, a behavior not seen in the Fit column.

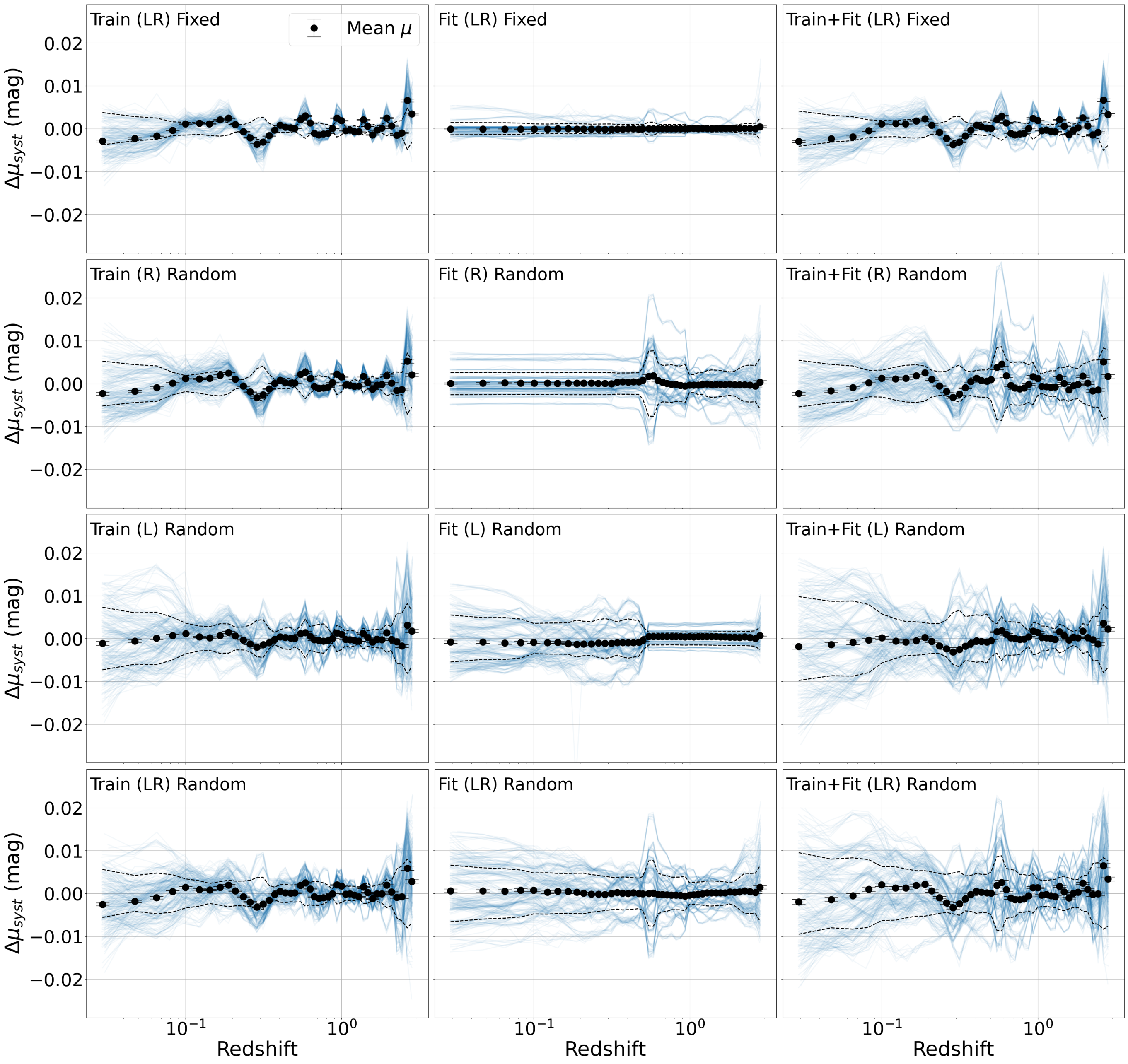

**Figure 6.** Redshift trend of distance modulus residuals relative to the nominal (no-systematics) case for all the analysis configurations. The $y$-axis label is as defined in Equation (12). In each panel, the light blue curves correspond to individual systematic realizations while the black points show the mean residual in redshift bins. The labels R, L, LR, Fixed, and Random are as defined in the beginning of this section. The top row uses the fixed zero-point and central wavelength shifts specified in the last row of Table 2, while all other rows apply Gaussian-distributed perturbations with zero mean and standard deviations given in that same table. Identical perturbation realizations are used within each row to enable direct comparison of training- vs. fitting-induced effects.

To test whether the systematic trends in $\Delta\mu_{\rm syst}$ are degenerate with cosmological evolution, we fit $\Delta\mu_{\rm syst}$ with a second-order polynomial in redshift and evaluate the resulting best-fit $\chi^2$. If $\Delta\mu_{\rm syst}(z)$ is well described by a smooth polynomial, it indicates that the systematic closely mimics cosmological evolution and hence cannot be self-calibrated in a cosmological fit using a covariance matrix; the expected impact is larger systematic uncertainty on cosmology parameters, and smaller $\mathcal{R}_{\rm FoM}$. For each panel in Figure 6, the resulting $\chi^2$ values are shown in Figure 7. We find that the Fit case yields very small $\chi^2$, consistent with our finding that Fit systematics result in smaller $\mathcal{R}_{\rm FoM}$ compared to Train.

As expected, the inclusion of systematics leads to weaker constraints as shown in Figure 8. When calibration systematics from Cal_SALT3 (Train, Fit) are included, the contours expand. The best-fit values of $w_0$ and $w_a$ vary slightly with the inclusion of systematics and are within $1\sigma$ of the simulated $\Lambda$CDM values.

Table 5 summarizes the degradation in cosmological constraining power, quantified by $\mathcal{R}_{\rm FoM}$ (Equation (11)), under different systematic uncertainty scenarios.

Across all analysis variants, introducing calibration systematics during training only (Train) leads to relatively modest degradation, with FoM reductions of order 7%–14%. When systematics are included in the fitting stage (Fit and Train, Fit), the impact is much greater, dropping the FoM to roughly 50% of the baseline value.

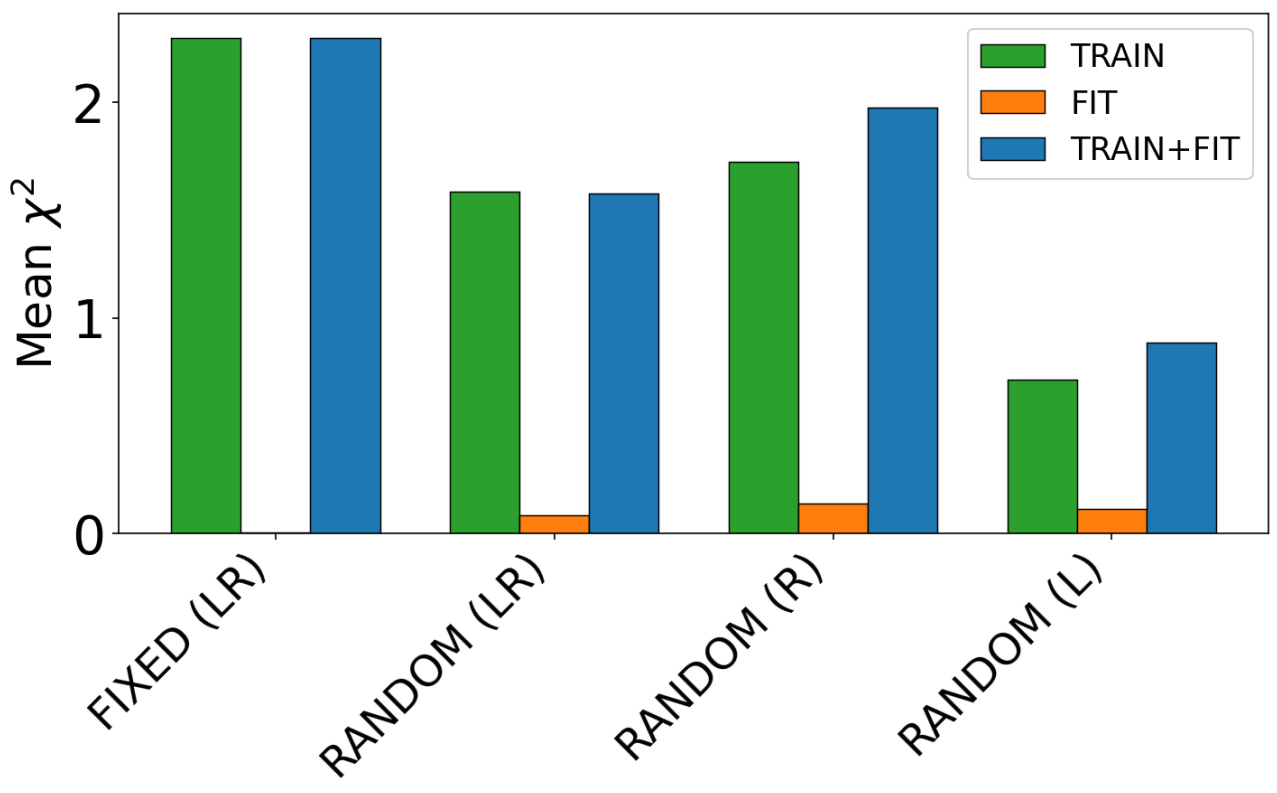

**Figure 7.** Mean $\chi^2$ from quadratic fits to $\Delta\mu_{\rm syst}$ among the analysis variants in Figure 6.

Figure 9 summarizes $\mathcal{R}_{\rm FoM}$ from fixed calibration perturbations in the individual filters of the two surveys. Among the Roman zero-point shifts, F106 and F129 are the most sensitive for the Fit and Train, Fit analysis variants, with up to 12% FoM degradation.

Among the Roman filter-wavelength shifts, the FoM degradation is less than 4%, and is much smaller compared to the degradation from zero-point shifts.

For Rubin-LSST filters, the Train, Fit analysis does not show any significant FoM drop in both zero-point offset and filter central wavelength shift. For the Fit analysis, the LSST $r$-band shows the most degradation, with about 19% FoM drop from zero-point offset. The LSST $i$-band shows a $\sim$7% FoM decrease in both the zero-point and filter central wavelength shift for the Fit analysis.

The FoM dependence on systematic shift amplitude is shown in Figure 10 for Train+Fit (LR) statistical-only and Cal_SALT3 calibration systematics for Train, Fit, and Train+Fit scenarios. A shift amplitude of 2, for example, means that the random numbers used to model the zero-point offsets and filter central wavelength shifts were drawn from a Gaussian of mean 0 and standard deviation of 2 mmag and 2 Å, respectively. We exclude the statistical-only FoMs for Train and Fit because their variations from the Train, Fit statistical-only FoM for the different shift amplitudes are not significant. The statistical-only FoMs across the different shift amplitudes remain stable across the shift amplitudes, confirming that the observed degradation arises from the imposed calibration perturbations.

FoM decreases monotonically with increasing shift amplitude across all calibration scenarios. If systematics are applied only during the training stage (Train), the resulting FoM degradation is significantly smaller compared to Fit and Train, Fit, indicating that the training procedure does not significantly contribute to error propagation through the pipeline. In contrast, the most severe loss of constraining power is from calibration systematics introduced in the fitting stage.

There are two notable features in Figure 10. First, the FoM-versus-shift relation is approximately linear apart from a fluctuation near shift value 6. Second, while one might naively expect Train, Fit to yield lower FoM than Fit we see the opposite trend for some shift values, particularly for shift $> 5$. In the paragraph below, we show that this crossover is a direct consequence of self-calibration (D. Brout et al. 2021), not a pipeline artifact. We have further confirmed that the crossover is statistically significant rather than a fluctuation. At shift amplitudes of 8, 9, and 10, the mean FoM difference between Train, Fit, and Fit is 21.9, 21.5, and 17.5, corresponding to significances of roughly $6\sigma$ to $8\sigma$.

The FoM impact of a systematic depends on how degenerate its $\Delta\mu_{\rm syst}$ shape is with the cosmological model. D. Brout et al. (2021) showed that the data can self-calibrate the component of a systematic that is orthogonal to the cosmological model, while the component degenerate with cosmology cannot be removed. In the Fit case, the calibration shift enters only at light-curve fitting and produces a smooth, redshift-dependent distance shift (the Fit column of Figure 6 shows no wiggles); this is nearly degenerate with cosmology and cannot be self-calibrated, so it degrades the FoM strongly. In the Train and Train, Fit cases, the shift is propagated through SALT3 training, distorting the model surfaces (Figure 5) resulting in a nonsmooth, nondegenerate structure in the residuals (the wiggles seen in the Train and Train, Fit columns of Figure 6, absent in Fit).

Moreover, the mean-$\chi^2$ diagnostic of Figure 11 supports this picture. At low amplitudes, Train and Train, Fit produce comparable $\chi^2$ while Fit remains near zero, consistent with well-conditioned fits in which the FoM differences between configurations are within statistical scatter. At large amplitudes, the training-induced, nonsmooth structure in Train and Train, Fit grows relative to the smooth Fit residuals. As shown in Figure C2, the scatter of the residuals grows with shift amplitude, and with it the self-calibratable component that drives the effect. Thus above shift amplitude $\sim$6, where this component is large enough to be effectively self-calibrated, Train, Fit recovers more constraining power than Fit and produces a higher FoM. The same reasoning explains why Fit and Train, Fit are statistically consistent at low amplitude rather than crossing over at all amplitudes: self-calibration is a large-signal effect, and at low amplitude the nondegenerate structure is too small relative to the statistical scatter for the covariance matrix to act on. We note that this self-calibration is preserved by our use of the REBIN scheme (Section 2.4), which bins in stretch and color as well as redshift; D. Brout et al. (2021) showed that binning in redshift alone removes the orthogonal information that enables self-calibration, whereas R. Kessler et al. (2023) showed REBIN closely recovers the unbinned results.

To verify that this behavior is not an artifact of the residual-fitting procedure used to produce Figure 11, we compute the mean $\chi^2$ with quadratic, cubic, or quartic polynomials, with a smoothing spline, and directly against zero after binning the residuals in different numbers of redshift bins (Appendix D); all yield the same qualitative trend. Furthermore, the calibration uncertainties expected for Stage IV surveys, $\sim$1–2 mmag (The LSST Dark Energy Science Collaboration et al. 2018), fall well within the low-amplitude regime where the FoM is well behaved.

Moreover, we characterize the dependence of the FoM on shift amplitude by fitting a straight line, $\langle {\rm FoM}\rangle = mA + b$, to each configuration in Figure 10 over the full amplitude range. The fitted slopes, in FoM per mmag, are $0.2 \pm 1.0$ (stat. only), $-11.0 \pm 0.7$ (Train), $-35.9 \pm 0.4$ (Fit), and $-32.7 \pm 0.4$ (Train, Fit). The stat-only slope is consistent with zero, confirming that the degradation is driven by the input systematics. The relation is not strictly linear above shift amplitude ∼6, but the linear approximation is reliable in the Stage IV regime. As a rule of thumb for setting calibration requirements, in the unmitigated Fit case a tolerance $\Delta\langle\text{FoM}\rangle$ corresponds to a maximum calibration error of $\sim\Delta\langle\text{FoM}\rangle/36$ mmag.

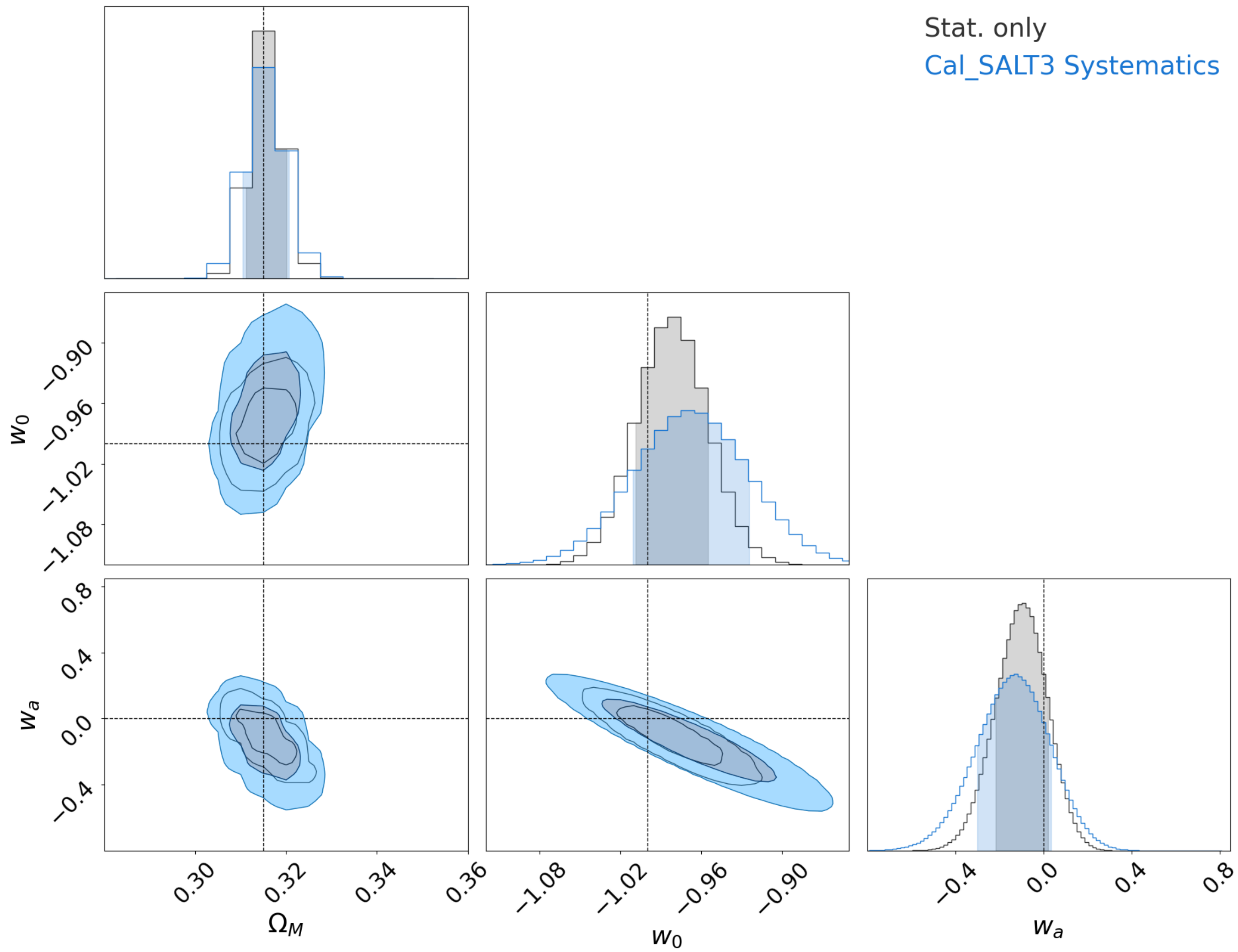

**Figure 8.** Contour plots for Flat$w_0w_a$CDM model from the simulation. We show statistical uncertainties and calibration systematics for Train+Fit(LR) analysis.

**Table 5**
Cal_SALT3 $\mathcal{R}_{\rm FoM}$ for the Different Analysis Variants

| Configuration | $\mathcal{R}_{\rm FoM}$ |
|---|---|
| TRAIN (LR) FIXED | 0.863 |
| FIT (LR) FIXED | 0.473 |
| TRAIN+FIT (LR) FIXED | 0.554 |
| TRAIN (R) RANDOM | 0.927 |
| FIT (R) RANDOM | 0.730 |
| TRAIN+FIT (R) RANDOM | 0.688 |
| TRAIN (L) RANDOM | 0.854 |
| FIT (L) RANDOM | 0.686 |
| TRAIN+FIT (L) RANDOM | 0.693 |
| TRAIN (LR) RANDOM | 0.868 |
| FIT (LR) RANDOM | 0.532 |
| TRAIN+FIT (LR) RANDOM | 0.485 |

Next, we compare ⟨FoM⟩ for the RANDOM (Figure 12 top left) and FIXED (Figure 12 bottom left) methods and find consistent results as expected with uncorrelated uncertainties. The right panels in Figure 12 compare the $w_0$ and $w_a$ uncertainties.

Finally, Figure 13 shows the impact on FoM from selectively excluding subsets. The most substantial FoM drop occurs if all SNe from the Rubin-LSST Deep Drilling Fields (DDF) are excluded. Excluding SNe from the Roman WIDE and DEEP fields results in comparable FoM reduction. These trends show the importance of each survey field in constraining cosmological parameter with Rubin-LSST DDF playing a critical role.

## 4. Conclusion

We analyzed simulated HLTDS + Rubin-LSST data to make significant progress in understanding and evaluating the impact of calibration systematics by isolating effects at the training and/or fitting stages, but several avenues remain open for extending this framework. While non-Ia contamination was included in the simulation, we simplified this study by assuming perfect knowledge of the host-galaxy redshifts, i.e., replacing the host photometric redshifts with their simulated spectroscopic redshifts for every event. This is also the first study to attempt SALT3 training using only NIR (prism) spectra, without the conventional set of optical spectra.

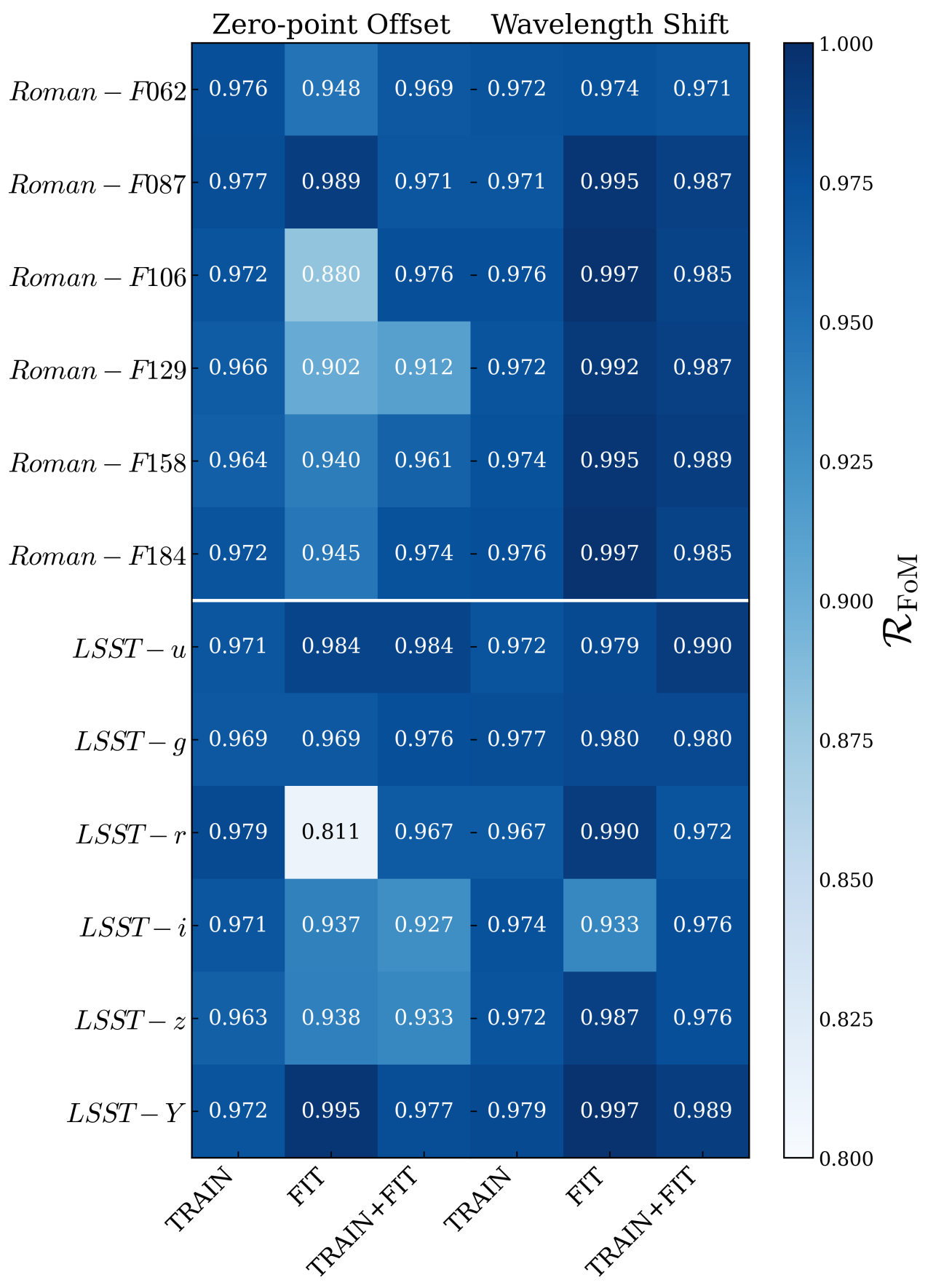

**Figure 9.** Cal_SALT3 $\mathcal{R}_{\rm FoM}$ showing the resulting degradation due to perturbations applied to individual filters. Fixed shifts of 5 mmag in zero-point and 5 Å in mean wavelength are applied. Train, Fit, and Train, Fit denote perturbations applied during the SALT3 training, light-curve fitting, or both stages, respectively. Left 3 columns are for zero-point shifts; right 3 columns are for filter-wavelength shifts.

Although the `SALT3.NIREXT` model used in the OpenUniverse2024 simulation (OpenUniverse et al. 2024) extends to 25,000 Å, enabling coverage into the near-infrared (NIR), we found subtle *SALTShaker* artifacts training with only prism spectra, and we therefore limited our rest-frame wavelength coverage only up to 11,000 Å. Another artifact of prism-only training is that the color dispersion is far smaller than in previous models, leading to a larger rejection of high-SNR (low-$z$) events. Future work is needed to understand these SALTShaker artifacts, and if both optical and NIR spectra are needed for optimal training results.

While this work provides new insights into the calibration systematics affecting SN cosmology, fully addressing real-world survey challenges will require incorporating other sources of uncertainties, testing a broader set of astrophysical models, and taking full advantage of the spectral range available in these next-generation observatories. Each of these improvements will be important for maximizing the scientific return of Rubin-LSST, Roman, and other Stage IV dark energy experiments.

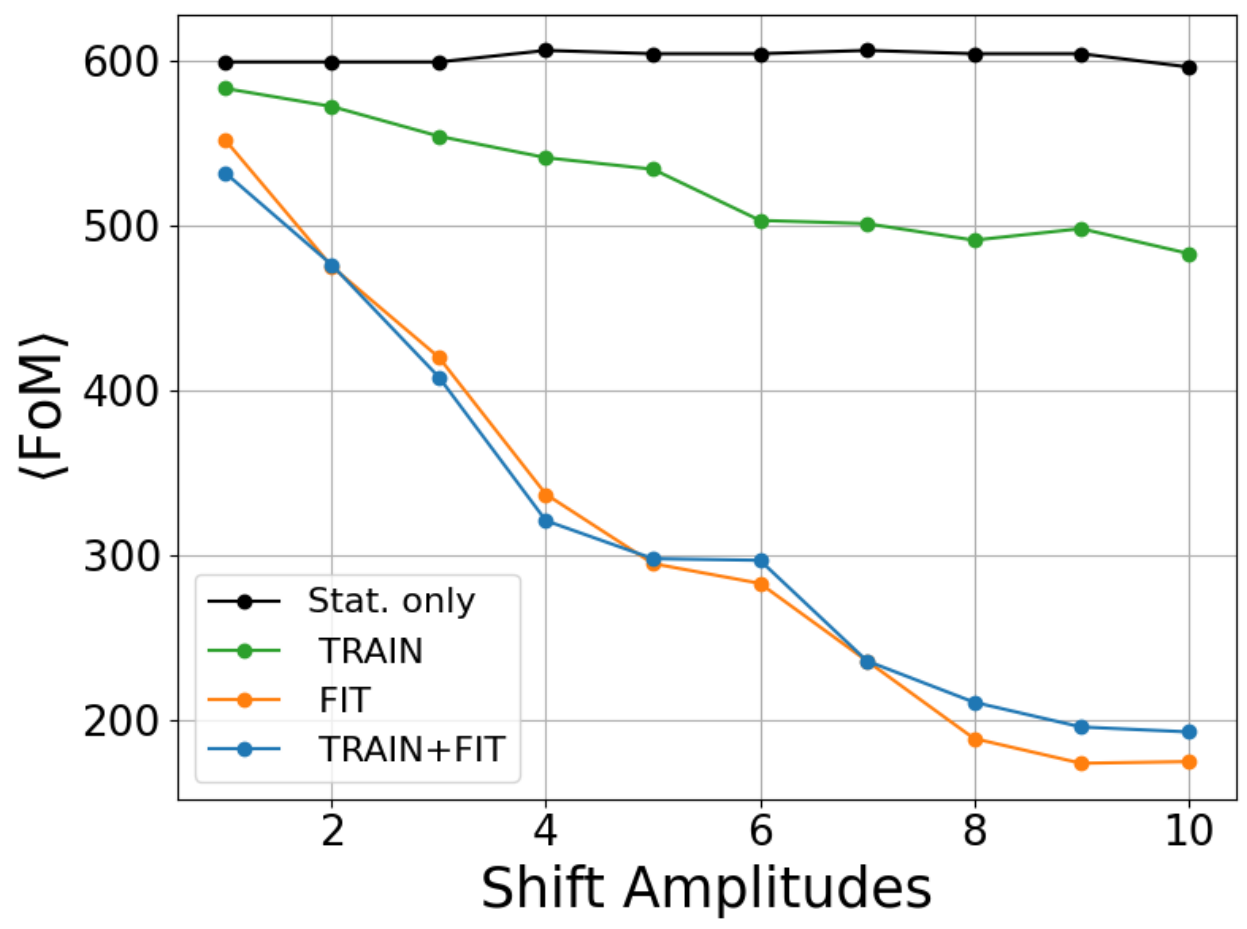

**Figure 10.** ⟨FoM⟩ vs. systematic shift amplitude. The statistical-only Train, Fit is shown in black, while Cal_SALT3 systematics are applied in Train (LR), Fit(LR), and Train, Fit(LR) analysis variants. A shift amplitude of 2 means that a zero-point offset of 2 mmag and filter mean wavelength of 2 Å were applied simultaneously.

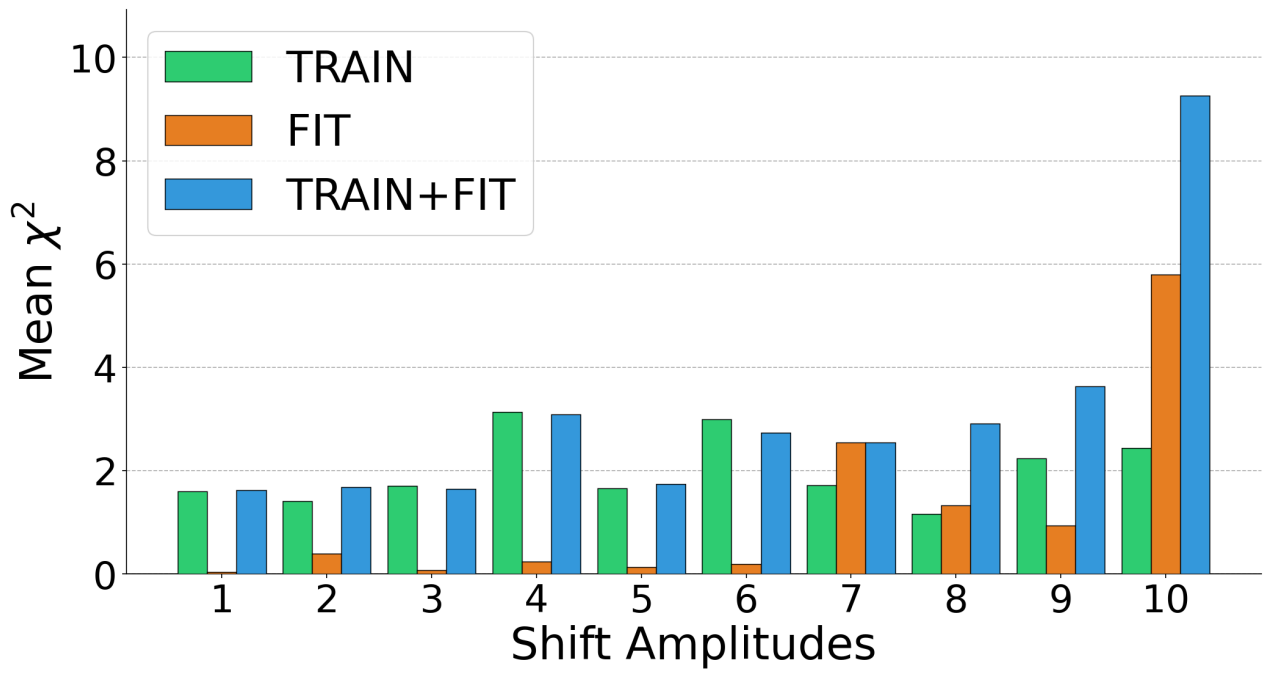

**Figure 11.** Mean $\chi$2 of $\Delta\mu_{\rm syst}$ (fitted against zero) vs. shift amplitudes, comparing TRAIN, FIT, and TRAIN+FIT for the LR Random configuration. A shift amplitude of 2 means that a zero-point offset of 2 mmag and filter mean wavelength of 2 Å were applied simultaneously.

We have quantified the impact of photometric calibration uncertainties on SN cosmology with next-generation surveys, propagating zero-point and filter-wavelength shifts through spectral model recovery, distance estimation, and dark energy constraints. We find that calibration errors—5 mmag in zero-points and 5 Å in filter mean wavelengths—reduce the dark energy FoM by ∼50% relative to the baseline statistical-only analysis.

Our finding that calibration systematics propagate primarily through the light-curve fitting stage rather than through model training differs from D. Brout et al. (2022b), who reported that changing the SALT2 surface and the light-curve zero-points simultaneously produces roughly twice the shift in the recovered $w$ as changing either alone, with the training-induced M0-surface change dominating the sensitivity. We note first that the two analyses measure different quantities: the $dw$ values in their Table 6 are central-value shifts computed from their single best-fit calibration solution, whereas we quantify the $\mathcal{R}_{\rm FoM}$ (a variance, marginalized through the systematic covariance matrix) under ensembles of zero-mean random calibration realizations. A coherent calibration offset produces a smooth redshift-dependent distance trend that shifts the central cosmology, and this is the effect their Table 6 captures; in our variance-based framework, by contrast, the training-induced distance residuals are not smooth in redshift and are therefore partially removed by the covariance matrix rather than biasing the recovered parameters. This is the same self-calibration behavior quantified in Figures 6, 7, and 11, and it naturally suppresses the training contribution to FoM degradation relative to the fitting contribution. Also, the number of SALT3 model realizations used differs—nine in D. Brout et al. (2022b) versus 30 here—and we show in Appendix B that the FoM depends significantly on this number, converging only by approximately 60 realizations. So a nine-realization estimate can differ substantially from the converged value. Further differences—the use of SALT3/SALTShaker rather than SALT2, prism-only training restricted to rest-frame wavelengths below 11,000 Å, and a substantially longer redshift baseline (0.025–3.0)—all act to change how a calibration shift is partitioned between smooth (cosmology-degenerate) and non-smooth (self-calibratable) distance trends, and hence between the training and fitting stages. The two analyses nonetheless agree that calibration is a dominant systematic that is only weakly reduced by self-calibration; the difference lies specifically in the relative importance of the training versus fitting stages, which is expected to be sensitive to these methodological choices.

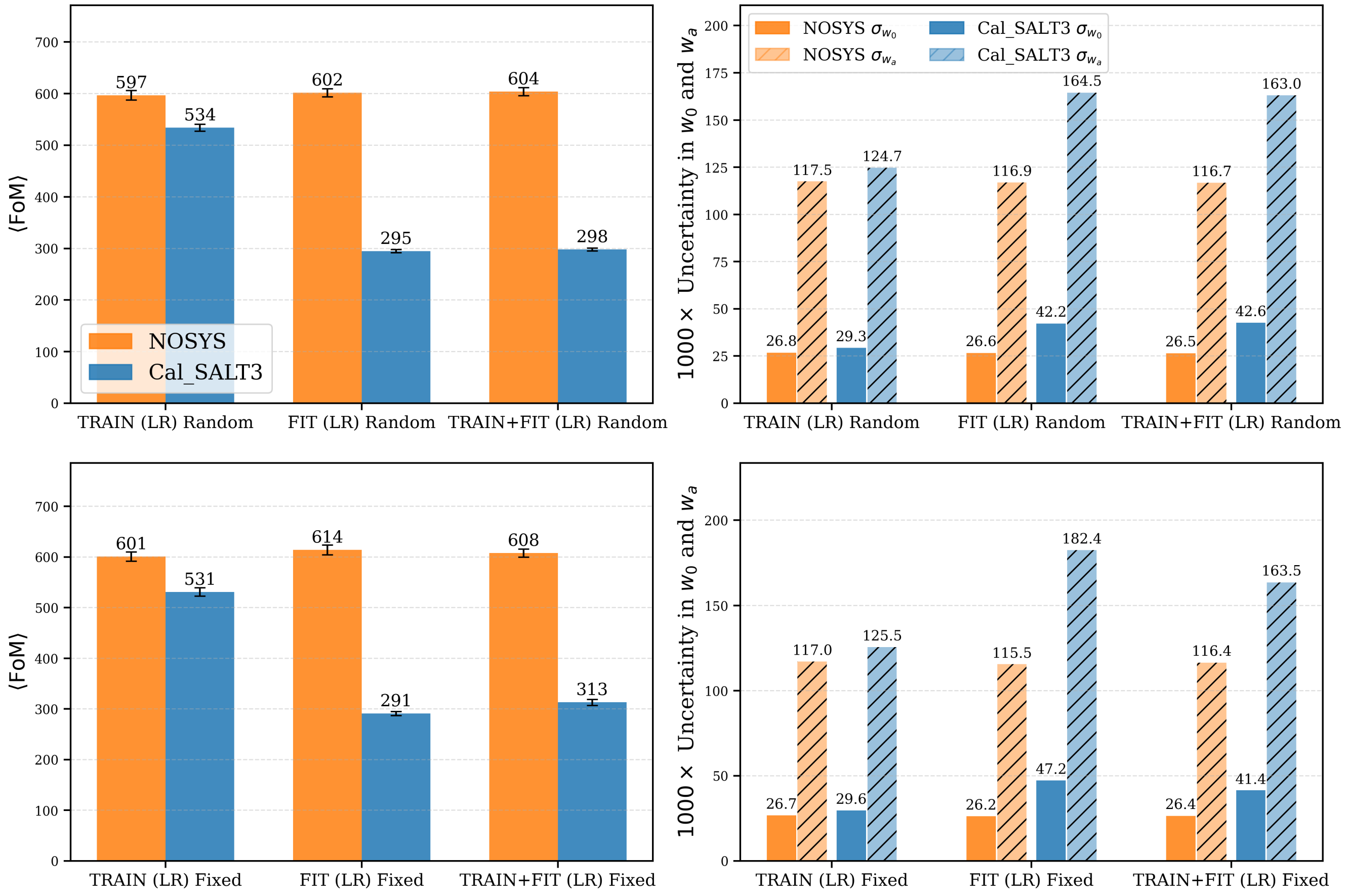

**Figure 12.** Comparison of ⟨FoM⟩ for RANDOM (upper left) and FIXED (bottom left) systematic shifts, and each panel shows TRAIN, FIT, TRAIN, FIT. The right panels show the corresponding average uncertainties on $w_0$ and $w_a$ multiplied by 1000, respectively.

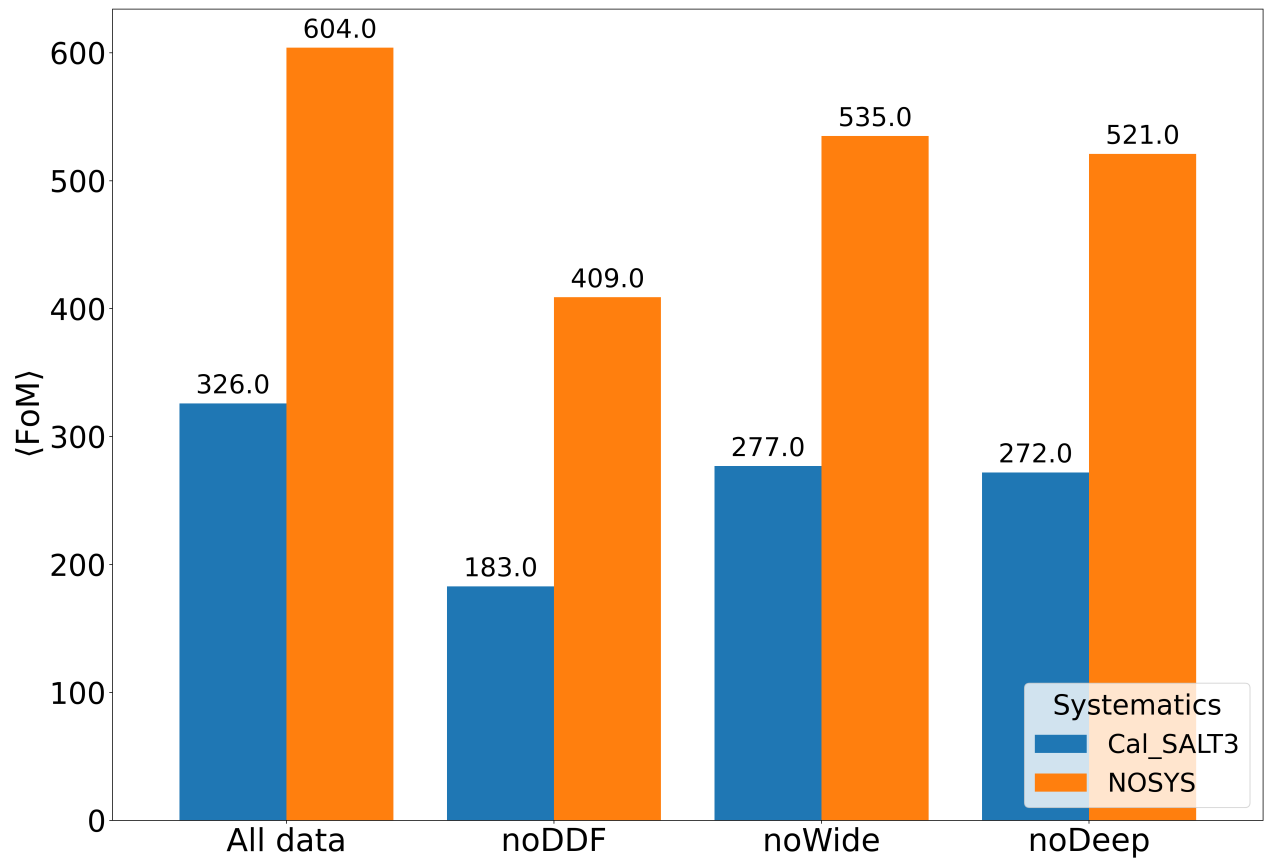

**Figure 13.** Impact of selectively excluding SNe from specific survey fields on the ⟨FoM⟩, derived from Train+Fit (LR) analysis. "noDDF" excludes all Rubin-LSST Deep Drilling Field SNe; "noWide" and "noDeep" exclude Roman Wide and Deep field SNe, respectively; "All data" retains the full sample with no exclusions. "NOSYS" denotes statistical uncertainties only, with no systematic errors included.

As these surveys prepare to deliver SN samples orders of magnitude larger than those available today, our results underscore that statistical gains will only translate into improved cosmological constraints if photometric calibration is controlled at the subpercent level. Achieving this will be essential to realizing the full dark energy science potential of the next generation of SN Ia cosmology programs.

## 5. Data and Software Availability

The configuration files, analysis scripts, and instructions required to reproduce the results of this work are publicly available at https://github.com/KeneAnumba/SALT3-model-Calibration-Systematics

## 6. Acknowledgments

K.A. and D.S. acknowledge support from NASA under contract 80NSSC24M0023 through the Roman Supernova Project Infrastructure Team, and from the University of Chicago's Research Computing Center for computing resources. D.O.J. acknowledges support from NSF grants AST-2407632, AST-2429450, and AST-2510993, NASA grant 80NSSC24M0023, and HST/JWST grants HST-GO-17128.028 and JWST-GO-05324.031, awarded by the Space Telescope Science Institute (STScI), which is operated by the Association of Universities for Research in Astronomy, Inc., for NASA, under contract NAS5-26555. This work is also funded in part by the Gordon and Betty Moore Foundation through Grant GBMF13900 to D.O.J. Furthermore, R.H. acknowledges that this material is based upon work supported by NASA under award number 80GSFC24M0006. Finally, J.P. thanks the LSST-DA Data Science Fellowship Program, which is funded by LSST-DA, the Brinson Foundation, the WoodNext Foundation, and the Research Corporation for Science Advancement Foundation; her participation in the program has benefited this work.

## Appendix A Validation of Trained Models

We compare both models' performance at recovering SN parameters for some simulated Roman light curves in Figure A1 below. Both models show broadly similar distributions for most of the key SN parameters and their uncertainties as well as redshift and redshift errors. This suggests general compatibility between the two models in terms of fitted light-curve parameters.

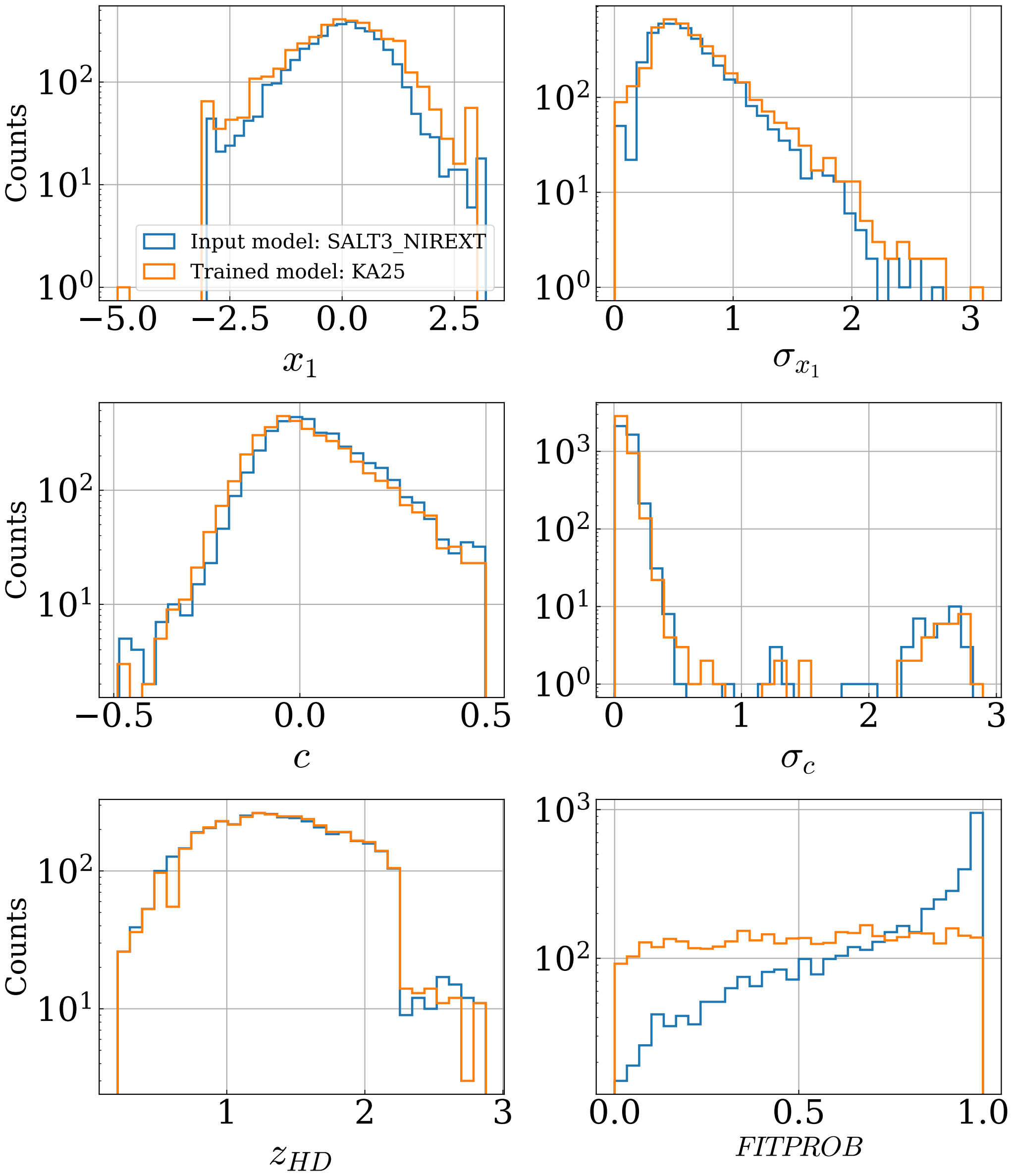

**Figure A1.** Comparison of fitted light-curve parameters and their associated uncertainties for the two models: the trained model—*KA25* and the input model—*SALT3_NIREXT*. Each panel shows the distribution of SN parameter for both models using the same binning. Distributions are plotted on a logarithmic *y*-axis to highlight differences across wide ranges. This comparison assesses how well the trained model recovers the underlying population of SN parameters.

## Appendix B
## Number of Model Realizations

The FoM reflects the constraining power on cosmological parameters, with higher values indicating tighter constraints. Figure B1 shows a strong dependence of FoM on the number of model realizations. Increasing the number of realizations systematically reduces the FoM from 354 in 4 realizations to about 270 in 60 realizations, where, we believe, convergence is achieved. We find that the FoMs remain stable across 60, 70, 80, 90, and 100 realizations. The relative contribution from the BINNED and REBIN components varies slightly as the number of realizations increases, but the overall stability indicates that cosmological constraints are well converged by at least ∼60 realizations; further increase in the number of realizations provides limited gains relative to the associated computational cost.

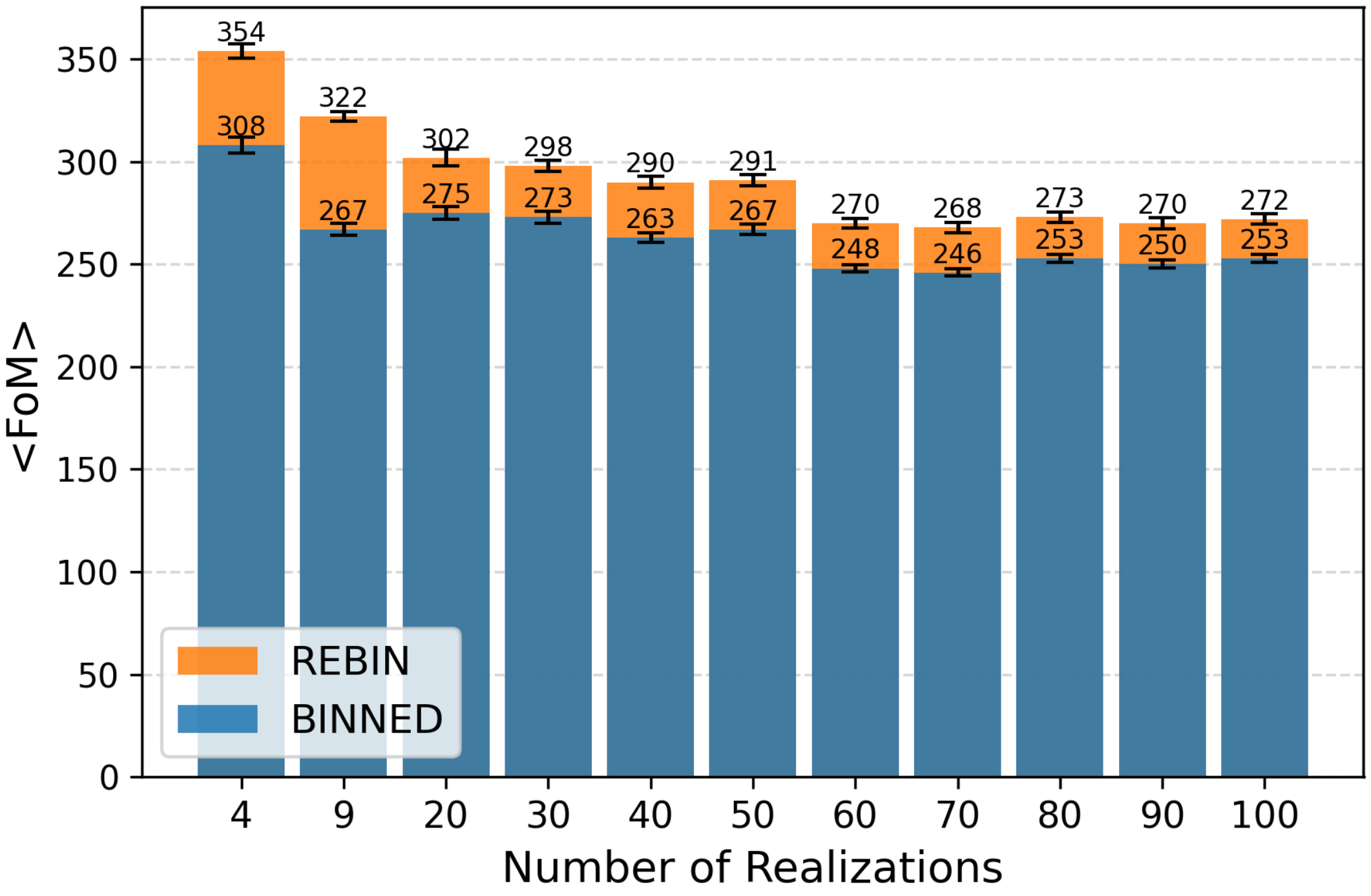

**Figure B1.** The dependence of FoM on the number model realizations used to sample the Cal_SALT3 calibration errors. Increasing the number of realizations improves the characterization of calibration uncertainties until convergence is reached around 60 model realizations, beyond which additional realizations yield diminishing returns relative to the computational cost. BINNED refers to the case where the Hubble diagram is binned in redshift space only while REBIN involves binning in redshift, stretch and color parameter spaces.

## Appendix C
## Different Shift Amplitudes

Furthermore, Figure C1 illustrates how varying shift amplitudes affect the recovered distances. As the shift amplitude increases, the variance grows correspondingly, showing an almost linear relationship in Figure C2.

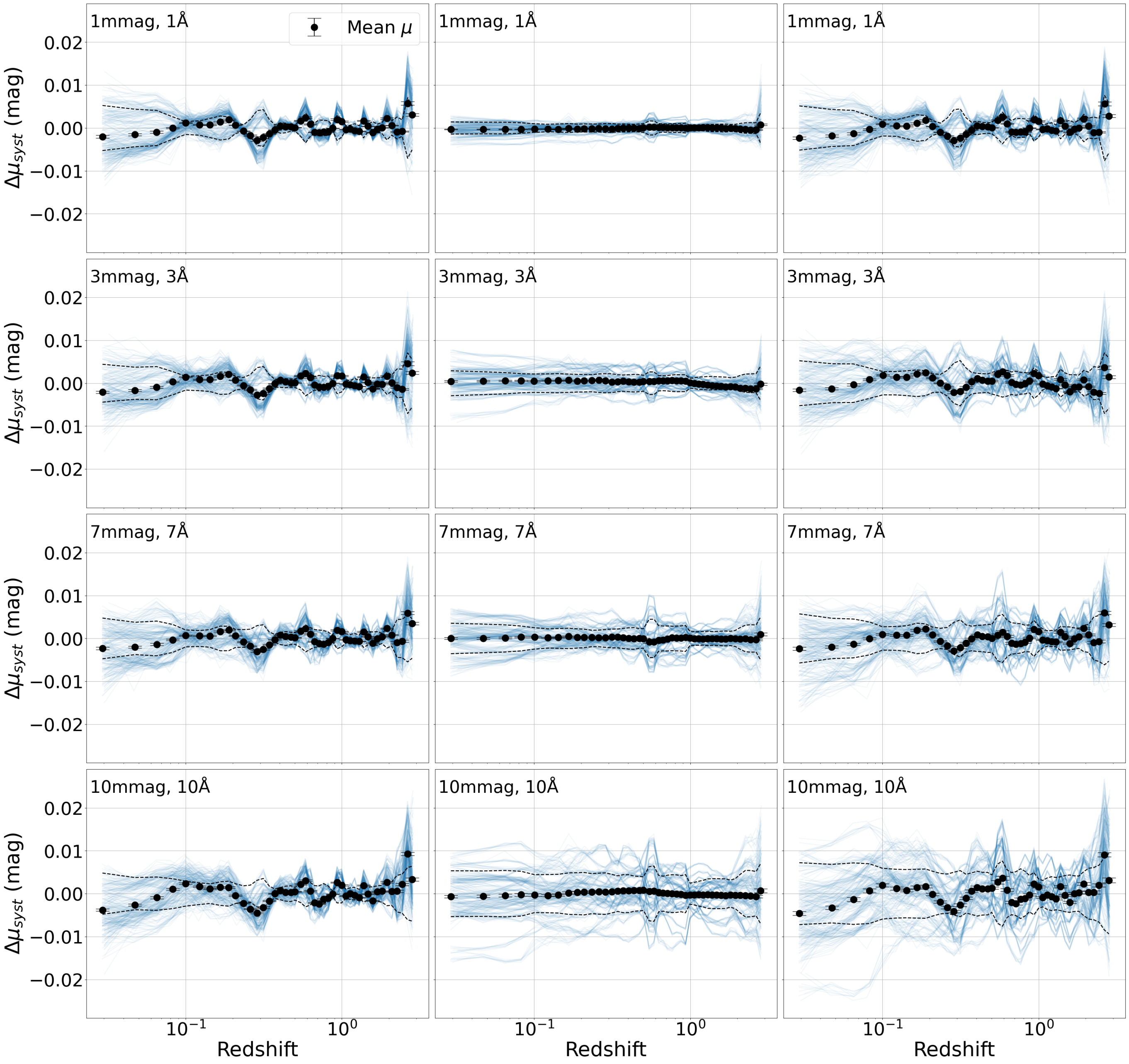

**Figure C1.** Redshift trend of distance modulus residuals relative to the nominal case for the different shift amplitudes. The shift amplitudes are annotated in the figures. The first, second, and third columns are Train, Fit, and Train, Fit, respectively.

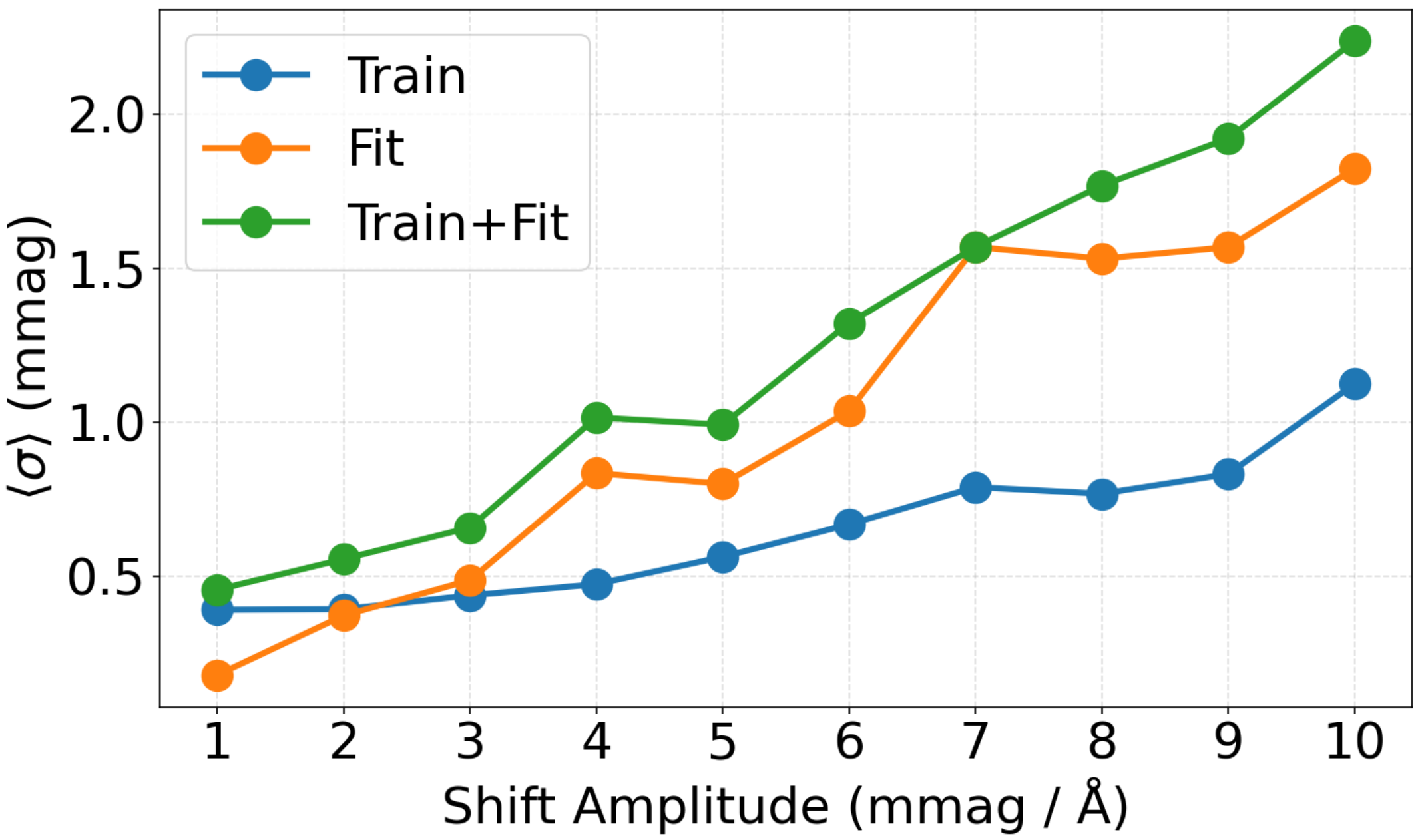

**Figure C2.** Mean standard error of the distance modulus shift as a function of calibration shift amplitude, for Train, Fit, and Train+Fit systematics. Each amplitude corresponds to equal zero-point and central wavelength shifts; for example, an amplitude of 2 corresponds to 2 mmag and 2 Å, respectively.

## Appendix D
## Robustness of the Mean $\chi2$ Trend to Fitting Method

To confirm that the trend in mean $\chi2$ versus amplitude shift (Figure 11), computed by comparing the redshift-averaged $\Delta\mu_{\rm syst}$ directly against zero, is not an artifact of that particular choice of statistic, we repeated the calculation using three alternative methods for characterizing the residual structure: (1) polynomial fits of increasing order, (2) a smoothing B-spline, and (3) direct binning in redshift. Allowing the polynomials to absorb progressively more redshift-dependent structure (2, 3, and 4 parameters) tests whether the trend seen in Figure 11 is being driven by a smooth trend that a polynomial curve could remove. Figure D1 shows that across all orders, the trend is preserved with only modest reductions as the polynomial order increases. Furthermore, as a nonparametric alternative to a fixed-order polynomial, we fit a smoothing cubic B-spline (weighted by the inverse of the $\Delta\mu_{\rm syst}$ uncertainties) to the same redshift-dependent residuals and computed $\chi2$ of the scatter around the spline (Figure D2). Because a spline can flexibly follow local structure that a low-order polynomial cannot, this serves as a check against the possibility that the polynomial trend above was masking real, nonpolynomial redshift dependence. Finally, we binned the residuals in redshift, using 10, 20, and 30 bins (Figure D3). This approach explores whether the effect is sensitive to specific choice of binning resolution. If the ordering and scaling of TRAIN, FIT, and TRAIN+FIT with amplitude shift is a genuine feature of the calibration systematic, rather than an artifact of how the “acceptable” baseline is defined, it should persist under all of these treatments.

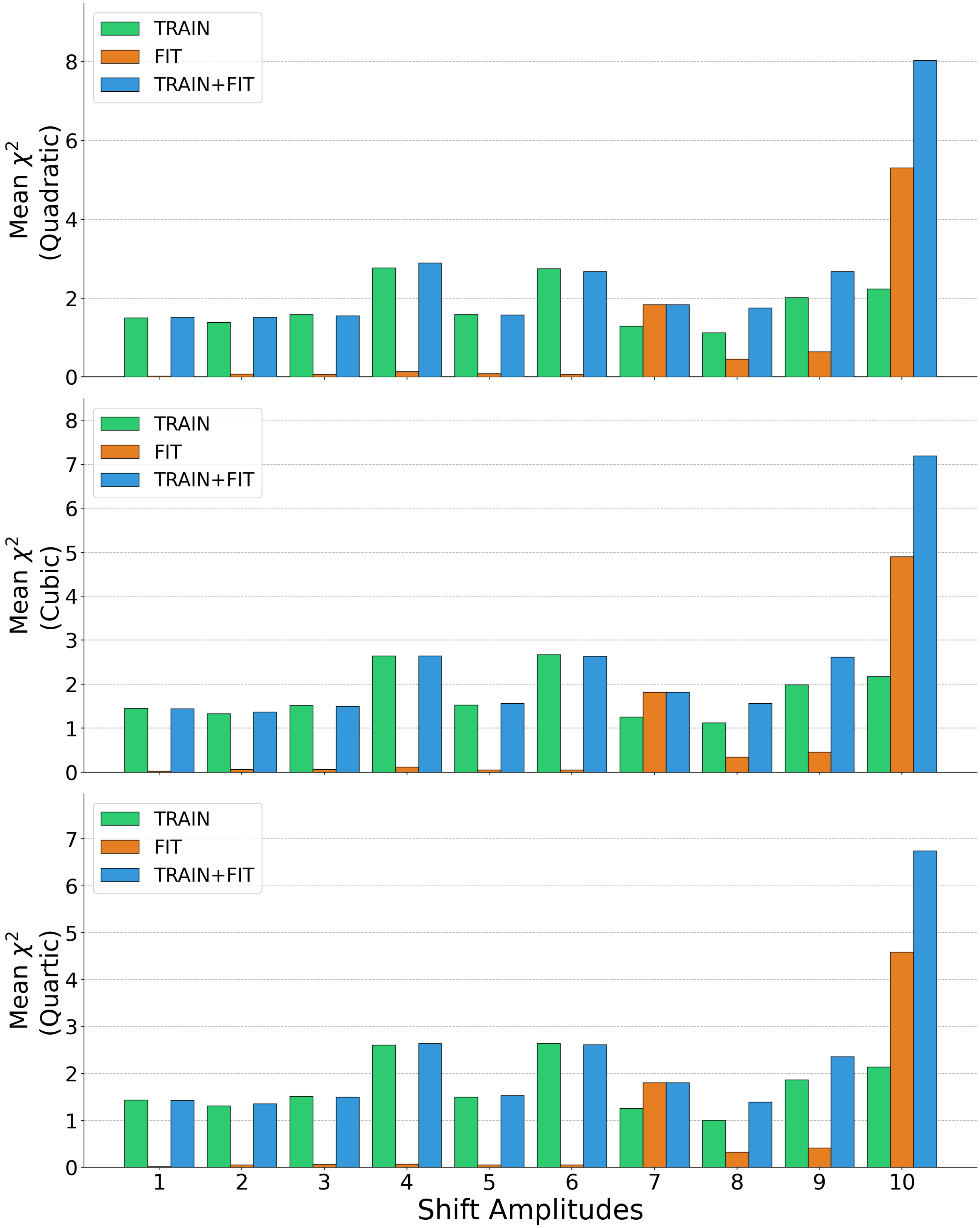

**Figure D1.** Mean $\chi 2$ of the $\Delta\mu_{\rm syst}$ relative to quadratic (top), cubic (middle), and quartic (bottom) polynomial fits.

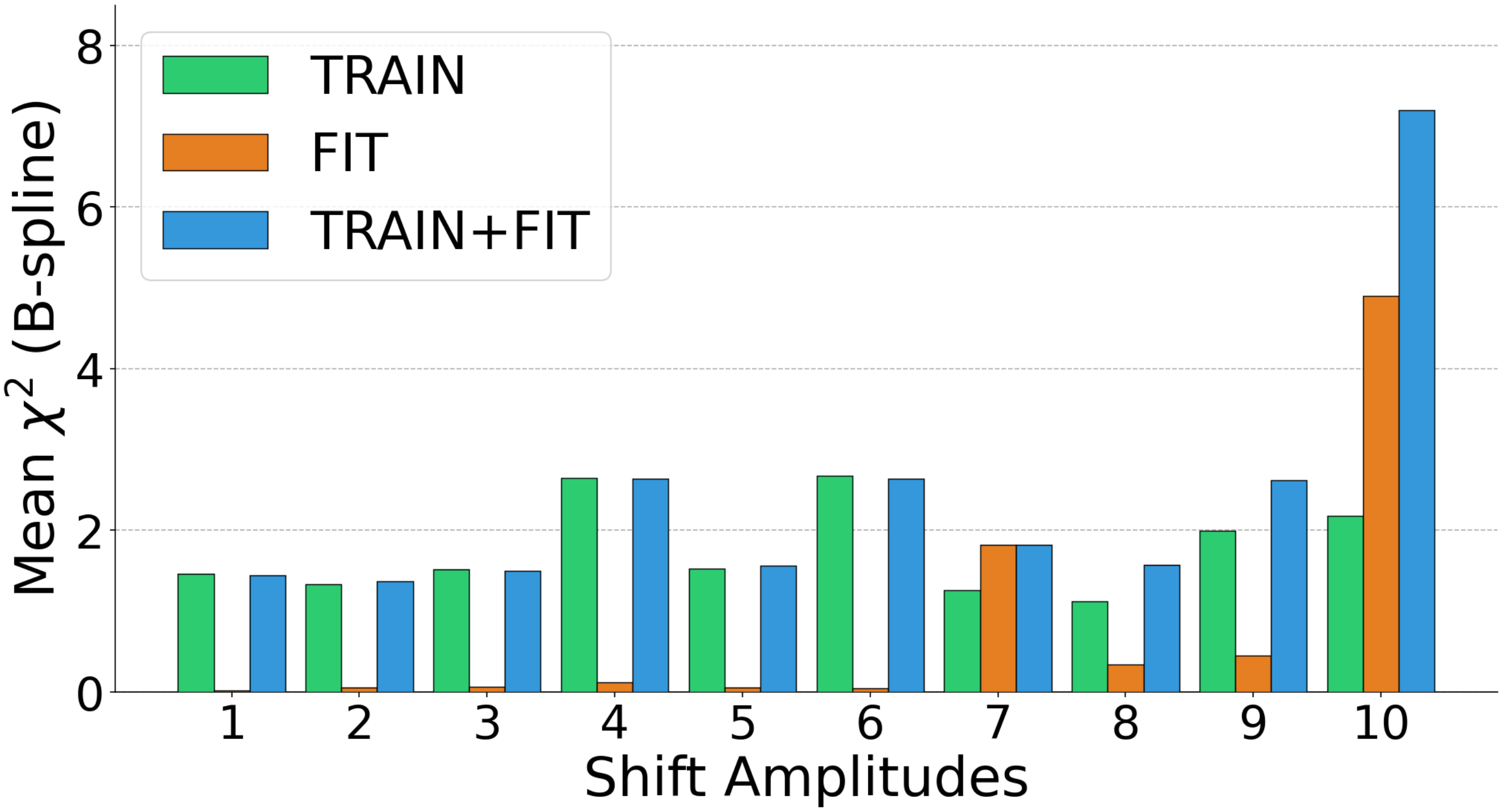

**Figure D2.** Mean $\chi 2$ of the $\Delta\mu_{\rm syst}$ relative to a smoothing cubic B-spline fit.

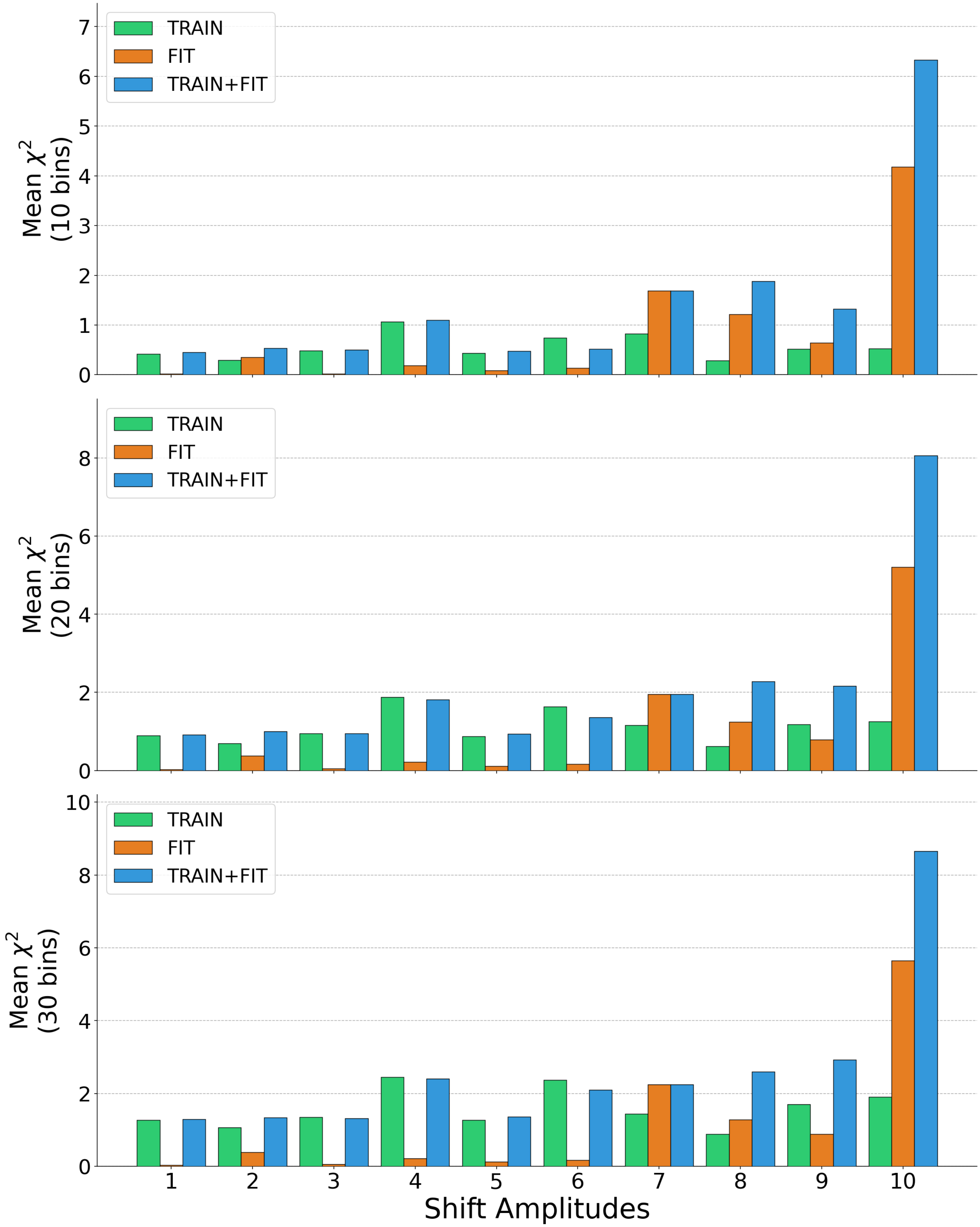

**Figure D3.** Mean $\chi 2$ of the $\Delta\mu_{\rm syst}$ against zero, computed in 10 (top), 20 (middle), and 30 (bottom) redshift bins.

## ORCID iDs

Kene Anumba https://orcid.org/0000-0001-8791-7273
David O. Jones https://orcid.org/0000-0002-6230-0151
Richard Kessler https://orcid.org/0000-0003-3221-0419
Daniel Scolnic https://orcid.org/0000-0002-4934-5849
W. D'Arcy Kenworthy https://orcid.org/0000-0002-5153-5983
Rebecca C. Chen https://orcid.org/0000-0003-3917-0966
Bastien Carreres https://orcid.org/0000-0002-7234-844X
Maria Vincenzi https://orcid.org/0000-0001-8788-1688
Erik R. Peterson https://orcid.org/0000-0001-8596-4746
Maria Acevedo https://orcid.org/0000-0002-5389-7961
Ben Rose https://orcid.org/0000-0002-1873-8973
Dillon Brout https://orcid.org/0000-0001-5201-8374
Jillian Paulin https://orcid.org/0000-0003-0408-366X
Rujuta A. Purohit https://orcid.org/0009-0003-6728-9291
Rebekah Hounsell https://orcid.org/0000-0002-0476-4206

## References

Abdul Karim, M., Aguilar, J., Ahlen, S., et al. 2025, PhRvD, 112, 083515
Adame, A., Aguilar, J., Ahlen, S., et al. 2025, JCAP, 2025, 021
Alam, S., Aubert, M., Avila, S., et al. 2021, PhRvD, 103, 083533
Albrecht, A., Bernstein, G., Cahn, R., et al. 2006, arXiv:astro-ph/0609591
Astier, P., Guy, J., Regnault, N., et al. 2006, A&A, 447, 31
Bailey, A. C., Vincenzi, M., Scolnic, D., et al. 2023, MNRAS, 524, 5432
Betoule, M., Kessler, R., Guy, J., et al. 2014, A&A, 568, A22
Brout, D., Hinton, S. R., & Scolnic, D. 2021, ApJL, 912, L26
Brout, D., Scolnic, D., Kessler, R., et al. 2019, ApJ, 874, 150
Brout, D., Scolnic, D., Popovic, B., et al. 2022a, ApJ, 938, 110
Brout, D., Taylor, G., Scolnic, D., et al. 2022b, ApJ, 938, 111
Chen, R., Scolnic, D., Rozo, E., et al. 2022, ApJ, 938, 62
Chen, R. C., Scolnic, D., Vincenzi, M., et al. 2025, MNRAS, 536, 1948
Chen, R. C., Guo, Z., Scolnic, D., et al. 2026, ApJ, 999, 81
Conley, A., Guy, J., Sullivan, M., et al. 2011, ApJS, 192, 1
Dai, M., Jones, D. O., Kenworthy, W. D., et al. 2023, ApJS, 267, 1
Dai, M., Kuhlmann, S., Wang, Y., & Kovacs, E. 2018, MNRAS, 477, 4142
Delgado, F., Cook, K., Miller, M., Allsman, R., & Pierfederici, F. 2006, SPIE, 6270, 62701D
Dilday, B., Kessler, R., Frieman, J. A., et al. 2008, ApJ, 682, 262
Guy, J., Astier, P., Nobili, S., Regnault, N., & Pain, R. 2005, A&A, 443, 781
Guy, J., Astier, P., Baumont, S., et al. 2007, A&A, 466, 11
Hamuy, M., Phillips, M. M., Suntzeff, N. B., et al. 1996, AJ, 112, 2391
Hinton, S., & Brout, D. 2020, JOSS, 5, 2122
Hlozek, R., Kunz, M., Bassett, B., et al. 2012, ApJ, 752, 79
HLTDS, H. 2025, arXiv:2505.10574
Hounsell, R., Scolnic, D., Foley, R. J., et al. 2018, ApJ, 867, 23
Ivezic, Z., Kahn, S. M., Tyson, J. A., et al. 2019, ApJ, 873, 111
Jones, D. O., Kenworthy, W. D., Dai, M., et al. 2023, ApJ, 951, 22
Jones, D. O., Scolnic, D. M., Riess, A. G., et al. 2018, ApJ, 857, 51
Jones, D. O., Scolnic, D. M., Foley, R. J., et al. 2019, ApJ, 881, 19
Joshi, B. A., Strolger, L.-G., Ryan, R. E., et al. 2022, ApJ, 941, 146
Kelly, P. L., Hicken, M., Burke, D. L., Mandel, K. S., & Kirshner, R. P. 2010, ApJ, 715, 743
Kenworthy, W. D., Jones, D. O., Dai, M., et al. 2021, ApJ, 923, 265
Kenworthy, W. D., Goobar, A., Jones, D. O., et al. 2025, A&A, 697, A125
Kessler, R., Hounsell, R., Joshi, B., et al. 2025, ApJ, 993, 116
Kessler, R., & Scolnic, D. 2017, ApJ, 836, 56
Kessler, R., Vincenzi, M., & Armstrong, P. 2023, ApJL, 952, L8
Kessler, R., Becker, A. C., Cinabro, D., et al. 2009a, ApJS, 185, 32
Kessler, R., Bernstein, J. P., Cinabro, D., et al. 2009b, PASP, 121, 1028
Kessler, R., Narayan, G., Avelino, A., et al. 2019a, PASP, 131, 094501
Kessler, R., Narayan, G., Avelino, A., et al. 2019b, PASP, 131, 094501
Kunz, M., Bassett, B. A., & Hlozek, R. A. 2007, PhRvD, 75, 103508
Lampeitl, H., Smith, M., Nichol, R. C., et al. 2010, ApJ, 722, 566
Lochner, M., Scolnic, D., Almoubayyed, H., et al. 2022, ApJS, 259, 58
LSST Science Collaboration, Abell, P. A., Allison, J., et al. 2009, arXiv:0912.0201
Marriner, J., Bernstein, J. P., Kessler, R., et al. 2011, ApJ, 740, 72
Mitra, A., Kessler, R., More, S., & Hlozek, R. 2023, ApJ, 944, 212
Mitra, A., Kessler, R., Chen, R. C., et al. 2026, ApJ, 1006, 122
OpenUniverse, The LSST Dark Energy Science Collaboration, The Roman HLIS Project Infrastructure Team, et al. 2024, arXiv:2501.05632
Perlmutter, S., Aldering, G., Goldhaber, G., et al. 1999, ApJ, 517, 565
Phillips, M. M. 1993, ApJL, 413, L105
Pierel, J. D. R., Jones, D. O., Dai, M., et al. 2021, ApJ, 911, 96
Pierel, J. D. R., Jones, D. O., Kenworthy, W. D., et al. 2022, ApJ, 939, 11
Planck Collaboration, Aghanim, N., Akrami, Y., et al. 2020, A&A, 641, A1
Popovic, B., Brout, D., Kessler, R., & Scolnic, D. 2023, ApJ, 945, 84
Popovic, B., Kenworthy, W. D., Ginolin, M., et al. 2026, A&A, 712, A131
Qu, H., Sako, M., Möller, A., & Doux, C. 2021, AJ, 162, 67
Qu, H., Sako, M., Möller, A., & Doux, C. 2021, AJ, 162, 67
Reuter, M. A., Cook, K. H., Delgado, F., Petry, C. E., & Ridgway, S. T. 2016, SPIE, 9911, 991125
Riess, A. G., Filippenko, A. V., Challis, P., et al. 1998, AJ, 116, 1009
Roman Observations Time Allocation CommitteeCore Community Survey Definition Committees 2025, arXiv:2505.10574
Rose, B. M., Baltay, C., Hounsell, R., et al. 2021, arXiv:2111.03081
Rose, B. M., Vincenzi, M., Hounsell, R., et al. 2025, ApJ, 988, 65
Rubin, D., Aldering, G., Betoule, M., et al. 2025, ApJ, 986, 231
Rubin, D., Aldering, G., Fruchter, A., et al. 2025, arXiv:2506.04327
Sánchez, B. O., Kessler, R., Scolnic, D., et al. 2022, ApJ, 934, 96
Schmidt, B. P., Suntzeff, N. B., Phillips, M. M., et al. 1998, ApJ, 507, 46
Scolnic, D., Rest, A., Riess, A., et al. 2014, ApJ, 795, 45
Scolnic, D. M., Jones, D. O., Rest, A., et al. 2018, ApJ, 859, 101
Spergel, D., Gehrels, N., Baltay, C., et al. 2015, arXiv:1503.03757
Sullivan, M., Conley, A., Howell, D. A., et al. 2010, MNRAS, 406, 782
Suzuki, N., Rubin, D., Lidman, C., et al. 2012, ApJ, 746, 85
Taylor, G., Lidman, C., Popovic, B., & Abbot, H. J. 2024, MNRAS, 528, 4643
Taylor, G., Jones, D. O., Popovic, B., et al. 2023, MNRAS, 520, 5209
The LSST Dark Energy Science Collaboration, Mandelbaum, R., Eifler, T., et al. 2018, arXiv:1809.01669
Tripp, R. 1998, A&A, 331, 815
Vincenzi, M., Brout, D., Armstrong, P., et al. 2024, ApJ, 975, 86
Wang, Q., Jones, D. O., Pierel, J. D. R., et al. 2025, arXiv:2512.25064
Wang, Y. 2008, PhRvD, 77, 123525